\documentclass[12pt,reqno]{amsart}
\usepackage{amssymb}
\usepackage{geometry}
\usepackage[latin1]{inputenc}
\usepackage[italian,english]{babel}
\usepackage{amsmath, amsfonts, amsthm}
\usepackage{subcaption}
\usepackage{graphicx}
\usepackage{pgfplots}
\usepackage{paralist}
\usepackage{mathtools, mathabx,bigints}
\numberwithin{equation}{section}
\usepackage{diagbox}
\usepackage{booktabs} % VM: Ambiente per seperatori tabelle
\usepackage{comment} % VM: Ambiente per commenti estesi
\usepackage{textgreek}
\usepackage{cleveref}
\usepackage{enumitem}
\usepackage{siunitx}
\usepackage{steinmetz}
\usepackage{adjustbox}
\usepackage{multirow}
\usepackage{url}
\usepackage[thinc]{esdiff}
\newtheorem{theorem}{Theorem}[section]

\newtheorem{cor}[theorem]{Corollary}
\newtheorem{proposition}[theorem]{Proposition}

\theoremstyle{definition}
\newtheorem{remark}[theorem]{Remark}
\theoremstyle{remark}

\newcommand{\z}{\boldsymbol{z}}

\newcommand{\re}[1]{\operatorname{Re}\{#1\}}
\newcommand{\im}[1]{\operatorname{Im}\{#1\}}

{\left\{\begin{array}{@{}l@{}}}{\end{array}\right.}
\patchcmd{\abstract}{\scshape\abstractname}{\textbf{\abstractname}}{}{}
\makeatletter %note a di pagina senza numero 1
\def\@makefnmark{} %note a di pagina senza numero 2
\makeatother %note a di pagina senza numero 3

\pgfplotsset{compat=1.18} 
\begin{document}

\title[Invariants in ECT via Dimensional Analysis]
{Invariants in Eddy Current Testing via Dimensional Analysis}
\author[Mottola, Sardellitti, Milano, Ferrigno, Laracca, Tamburrino]{Vincenzo Mottola$^{1,2,3}$, Alessandro Sardellitti$^4$, Filippo Milano$^{1,2}$, Luigi Ferrigno$^{1,2}$, 
Marco Laracca$^5$, Antonello Tamburrino$^{1,2,3}$}\footnote{\tiny\\$^1$Department of Electrical and Information Engineering, University of Cassino and Southern Lazio, 03043 Cassino (FR), Italy \\
$^2$European University of Technology, European Union, Cassino, 03043, Italy.\\
$^3$EUT+ Institute of Nanomaterials and Nanotechnologies-EUTINN, European University of Technology, European Union, Cassino, 03043, Italy.\\
$^4$Faculty of Technological and Innovation Sciences, Universitas Mercatorum, 00186 Rome (RM), Italy \\ 
$^5$Department of Astronautics, Electrical and Energy Engineering, Sapienza University of Rome, 00186 Rome (RM), Italy \\ 
Email: vincenzo.mottola@unicas.it, alessandro.sardellitti@unimercatorum.it, filippo.milano@unicas.it, luigi.ferrigno@unicas.it, marco.laracca@uniroma1.it, antonello.tamburrino@unicas.it (corresponding author).}

\setcounter{tocdepth}{1}

\begin{abstract}
The Buckingham's \textpi \, theorem has been recently introduced in the context of Non destructive Testing \& Evaluation (NdT\&E) , giving a theoretical basis for developing simple but effective methods for multi-parameter estimation via \emph{dimensional analysis}. Dimensional groups, or \textpi-groups, allow for the reduction of the number of parameters affecting the dimensionless measured quantities.

In many real-world applications, the main interest is in estimating only a subset of the variables affecting the measurements. An example is estimating the thickness and electrical conductivity of a plate from Eddy Current Testing data, regardless of the lift-off of the probe, which may be either uncertain and/or variable. Alternatively, one may seek to estimate thickness and lift-off while neglecting the influence of the electrical conductivity, or to estimate the electrical conductivity and the lift-off, neglecting the thickness.

This is where the concept of invariants becomes crucial. An invariant transformation is a mathematical mapping or a specific operating condition that makes the measured signal independent of one or more of these uncertain parameters. Invariant transformations provide a way to isolate useful signals from uncertain ones, improving the accuracy and reliability of the NdT results.

The main contribution of this paper is a systematic method to derive \emph{invariant} transformations for frequency domain Eddy Current Testing data, via dimensional analysis. The proposed method is compatible with real-time and in-line operations. 

After its theoretical foundation is introduced, the method is validated by means of experimental data, with reference to configurations consisting of plates with different thicknesses, electrical conductivity, and lift-off. The experimental validation proves the effectiveness of the method in achieving excellent accuracy on a wide range of parameters of interest.

\end{abstract}
\maketitle

\section{Introduction}
The advent of the Industry 4.0 and NDE 4.0 era has significantly increased the demand for advanced Non Destructive Testing (NDT) techniques that can be integrated in automated and in-line inspection systems \cite{NDE40,nde40_2,ZDM,Corr}. Among the various NDT methods, Eddy Current Testing (ECT) is particularly promising for the inspection of metallic components in the framework of the Industry 4.0 paradigm, due to its contactless nature, allowing high-speed and fully automated inspections, unlike other methods such as ultrasonic testing, where coupling media are generally required \cite{ultra2}. In addition, ECT systems are cheaper if compared to other inspection methods, which require expensive equipment and/or controlled environments \cite{ect_rev}.

The ECT of metallic plates is a well-studied topic and several methods are available to estimate the electrical conductivity of the plate \cite{cond1,cond2,cond3,cond4}, its thickness \cite{thick1,thick2,thick3,thick4}, or magnetic permeability \cite{perm1,perm2,perm3}, and lift-off \cite{Lift_off_1,Lift_off_2}. All these quantities are directly related to the state of health of the material and of its non-conductive coating, if any \cite{app1,app2,app3,app4}. The possibility to integrate eddy current probe in in-line operation allows, then, to evaluate the product quality during their fabrication.

One of the main concerns is that the signals produced during an Non destructive Testing \& Evaluation (NdT\&E) test can be influenced by several factors that are not related to the quantities of interest. These factors are often referred to as nuisance parameters, to indicate that they can differ across measurements or may have uncertain, or even unknown, values. The most common of these is the lift-off, which is the distance between the probe and the specimen. A slight change in lift-off can cause a significant variation in the eddy current signal, altering or masking the useful signal from the quantities of interest, thus leading to a false indication. In practice, however, other parameters may also act as nuisance parameters, such as the thickness of an irregular or corroded plate, or a non-uniform distribution of electrical conductivity caused by non-idealities in the fabrication process.
This is where the concept of invariants becomes crucial. An invariant is a mathematical transformation or a specific operating condition that makes the measured signal independent of one or more of these nuisance parameters, such as lift-off, electrical conductivity or thickness. Identifying and using invariants, it is possible to isolate useful signals from nuisance ones, improving the accuracy and reliability of the test.

For Pulsed ECT, lift-off point of intersection (LOI) based methods have been proposed to overcome such difficulty \cite{LOI1, LOI2, LOI3}. These methods rely on the LOI, which is the time where the transient response of the coil is independent of lift-off variations and, in turn, can be directly related to specific material properties or geometric features of the specimen. In \cite{Norm1}, a different strategy is proposed employing a two-stage method, where two different reference signals are employed and the lift-off is reduced via proper minimization. Other strategies include the use of artificial neural networks \cite{ANN} and a proper optimization of the coil design \cite{PD1,PD2,PD3}.

In the frequency domain, different lift-off invariant features have been identified. For nonmagnetic plates, the phase of the measured impedance \cite{PH1,PH2} is governed mainly by thickness and electrical conductivity, remaining approximately unaffected by lift-off. Similarly, the (compensated) peak frequency of the inductance change \cite{COMP1} shows minimal dependence on lift-off. For magnetic specimens, a related lift-off invariance phenomenon \cite{LII} has been reported, i.e. a specific frequency exists at which the inductance change is almost the same for any lift-off. Finally, a strategy based on a lift-off invariant transformation of the measured data is proposed in \cite{TR1}.

Furthermore, related invariance effects have been identified, such as the electrical conductivity invariance lift-offs phenomenon \cite{CIP}, in which, at a specific lift-off and fixed frequency, there exists a point where the inductance is nearly insensitive to changes in electrical conductivity and depends solely on the magnetic permeability. In \cite{PIP}, the dual phenomenon, i.e., the permeability invariance phenomenon is presented for the estimation of the electrical conductivity in the presence of uncertainties on the magnetic permeability. In \cite{IT_AU}, an invariance transformation procedure is introduced able to make the signal insensitive to the magnetic permeability of the specimen, while retaining the information about the presence of defects.

This paper falls into this framework of development of invariant transformations, immune to nuisance parameter(s). Specifically, it proposes a general methodology for generating systematically invariant transformations via proper projections of level surfaces. The proposed method will be described with application to the simultaneous estimate of: (i) the electrical conductivity and the thickness of a metallic specimen, when the lift-off is unknown (lift-off invariance); (ii) the electrical conductivity and the lift-off, when the thickness is unknown (thickness invariance); and (iii) the thickness and the lift-off, when the electrical conductivity is unknown (electrical conductivity invariance).

The derivation of these invariant features is based on a new general paradigm for the study of NdT\&E problems \cite{Buck_main}, based on dimensional analysis and the application of Buckingham $\pi$-theorem \cite{Buck1}.
Dimensional analysis and Buckingham $\pi$-theorem allow us to systematically reduce the number of variables required to describe a physical problem by rewriting the original equations in a dimensionless form.
There are several advantages that arise from this approach. From the one hand, the reduction of the number of variables enables us to develop new methods, possibly compatible with real-time applications. On the other hand, the way the dimensionless quantities are formed highlights nontrivial relationships between the physical parameters. 

Most of the available methods on invariant transformation for the suppression of lift-off noise are based on proper approximations of the analytical model proposed by Dodd and Deeds \cite{Dodd_deeds}. This limits the range of applicability to the canonical configuration and produces approximate invariant transformations. In contrast, the approach proposed in this work (i) allows us to systematically identify quantities that are \emph{exactly} invariant, since no approximations are needed, and (ii) to treat arbitrary geometries, since no assumptions are required on the configuration of the coils and/or of the specimen.

The paper is organized as follows: in Section \ref{sec:foundation} the application of Buckingham $\pi$-theorem is briefly revised, in Section \ref{sec:inter} the new proposed methods are developed, in Section \ref{sec:exp} some experimental examples of application are shown and, finally, in Section \ref{sec:conclusion} the conclusions are drawn.

\section{Dimensionless description of the ECT problem}\label{sec:foundation}
Without loss of generality, a single coil configuration is considered, where the same coil generates the excitation field and senses the total magnetic flux density. It is worth noticing that for other coil configurations the following reasoning can be applied with marginal changes.

The measured quantity for this configuration is the impedance ($\dot{Z}$) of the coil, i.e. the ratio between the complex phasors of the induced voltage and the driving current applied to the sensing coil. In typical ECT operations, to enhance the performances of a method, the difference between the impedance measured in the presence of the plate ($\dot{Z}_{plate}$) and in the absence of the plate ($\dot{Z}_{air})$ is processed. The quantity of interest is, therefore, $\Delta \dot{Z}=\dot{Z}_{plate}-\dot{Z}_{air}$.

\begin{figure}[htb]
\centering
\includegraphics[width=.9\linewidth]{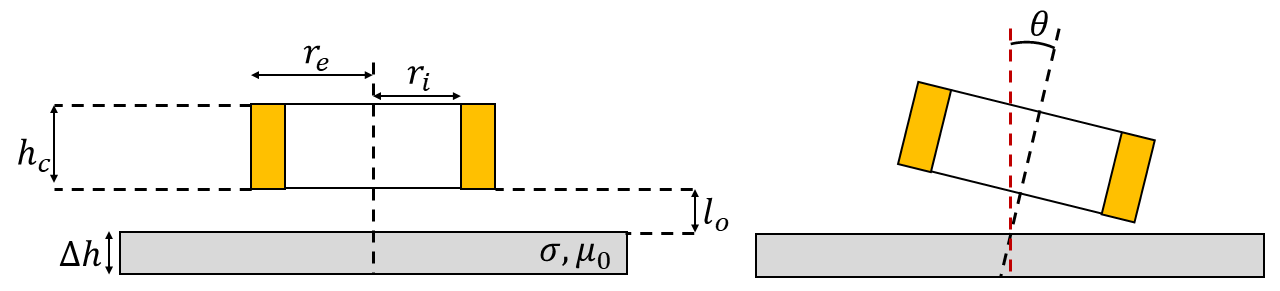}
\caption{Geometry of the problem. A single coil configuration is considered. The metallic plate is depicted in gray, while the coil is in yellow. The coil is placed at a distance $l_o$ and tilted by $\theta$ from the metallic plate.}
\label{fig:coils}
\end{figure}

In an ECT experiment, several different quantities, ranging from the geometry of the probe to the electromagnetic properties of the plate, determines $\Delta\dot{Z}$. The dependence of the measured quantity on the influencing parameters can be summarized as follows
\begin{equation}\label{eqn:pmod}
    \frac{\Delta\dot{Z}}{N^2}=\overline{f}(\omega,\sigma,\nu_0,\Delta h,l_o,D,\mathbf{t},\theta),
\end{equation}
where $\sigma$ and $\Delta h$ are the electrical conductivity, and thickness of the non-magnetic plate, respectively, $\nu_0$ is the free-space magnetic permeability, $D=r_e$ is the characteristic dimension of the probe, $\mathbf{t}=\left(r_i/D,h_c/D \right)$ is a vector containing the remaining (normalized) geometrical parameters of the coil, $l_o$ is the distance between the probe and the plate (lift-off) and $\theta$ is the tilting angle  (see Figure \ref{fig:coils}), $\omega$ is the angular frequency of the driving current, and $N$ is the number of turns of the coil.

Equation \eqref{eqn:pmod} involves ten different physical quantities: eight real scalars, one real vector, and one complex scalar.

In parameter estimation, dimensional analysis represents an extremely powerful tool, as proposed in \cite{Buck_main}. Indeed, Buckingham \textpi \ theorem allows for a systematic and significant reduction in the number of independent variables determining the measured quantity, thus diminishing the complexity of the analysis.

%The high number of variables makes the ECT problem difficult to treat, given the nonlinear nature of the function $f$, also. In \cite{Buck_main}, a new direction to deal with the ECT problem has been introduced, based on dimensional analysis. These techniques, and, in particular, the Buckingham's \textpi\,theorem, allow for a systematic reduction of the number of variables needed to describe the problem, thus diminishing the complexity of the analysis. For the sake of clarity, the main result from \cite{Buck_main}, it is briefly described below.

As proved in \cite{Buck_main}, \eqref{eqn:pmod} can be recast in a dimensionless form as
\begin{equation}\label{eqn:amod}
    \frac{\Delta \dot{Z} \nu_0}{\omega D N^2} = \overline{F} \left( \frac{D}{\delta},\frac{\Delta h}{D},\frac{l_0}{D},\mathbf{t},\theta \right),
\end{equation}
where $\delta=\sqrt{\frac{2\nu_0}{\omega\sigma}}$ is the skin-depth. In practical applications the parameters $\mathbf{t}$ and $\theta$ are known. Under this assumption, \eqref{eqn:amod} reduces to
\begin{equation}
\label{eqn:amod1}
    \frac{\Delta \dot{Z} \nu_0}{\omega D N^2 } = \overline{F}_p \left( \frac{D}{\delta},\frac{\Delta h}{D},\frac{l_0}{D}\right),
\end{equation}
while the corresponding dimensional equation is
\begin{equation}\label{eqn:pmod1}
    \frac{\Delta\dot{Z}}{N^2}=\overline{f}_p(\omega,\sigma,\nu_0,\Delta h,l_o,D).
\end{equation}
By comparing Equation \eqref{eqn:pmod1} and \eqref{eqn:amod1}, the impact of Buckingham \textpi-theorem is evident. Indeed, Equation \eqref{eqn:pmod1} that involves six different independent variables can be reduced to the equivalent form of \eqref{eqn:amod1}, where only three independent variables appear. This is very relevant because the reduction in the dimension of the space of the unknown variables determines a reduction in the complexity of parameter estimation problems, leaving room for the development of new real-time inversion methods.

\begin{remark}
    It is worth noticing that Equations \eqref{eqn:amod} and \eqref{eqn:amod1} are derived without any approximations.
\end{remark}

Hereafter, the following notation for the dimensionless quantities appearing in Equation \eqref{eqn:amod1}, called \textpi-groups,  is adopted
\begin{equation}\label{eqn:pigroups}
    \overline{\pi}_1=\frac{\Delta\dot{Z}\nu_0}{N^2\omega D},\quad \pi_2=\frac{D}{\delta},\quad \pi_3=\frac{\Delta h}{D} \quad \pi_4=\frac{l_o}{D}
\end{equation}

\section{Invariant quantities in ECT}\label{sec:inter}
The aim of this section is to introduce a method to derive invariant transformation via dimensional analysis.

A non-magnetic plate tested with ECT in the frequency domain is the configuration of interest to present the proposed method. A set of invariant quantities in ECT is derived, which allows the estimation of some parameters of interest, without the knowledge of all the quantities needed to describe the problem.

Specifically, it is shown that: (i) electrical conductivity and plate thickness can be estimated when lift-off acts as a nuisance parameter; (ii) thickness and lift-off can be estimated when electrical conductivity is treated as a nuisance parameter; and (iii) electrical conductivity and lift-off can be estimated when plate thickness is considered as a nuisance parameter.

Hereafter, it is assumed that $\mathbf{t}$ and $\theta$ are known, that is, the data are modeled by \eqref{eqn:amod1}.

\subsection{Representation of the data}

The derivation of such invariant quantities is carried out starting from Equation \eqref{eqn:amod1}. The dimensionless model is convenient, since it allows for a representation of the problem in a three dimensional space. Indeed, $\overline{\pi}_1$, the dimensionless version of the complex measured data, depends solely on the real independent variables $\pi_2$, $\pi_3$ and $\pi_4$. 

\subsubsection{Level surfaces}
Let $\mathbb{P}=(0, +\infty) \times (0, +\infty) \times [0, +\infty)$ be the parameter space for the triple $(\pi_2,\pi_3,\pi_4)$. A natural and powerful choice to represent the data is by means of the so-called level surfaces, which arise from the dimensionless measured data $\overline{\pi}_1$. Specifically, a family of level surfaces can be introduced in $\mathbb{P}$ as $g(\overline{F}_p(\pi_2,\pi_3,\pi_4))=c$, where $g: \mathbb{C} \mapsto \mathbb{R}$ and $c \in \mathbb{R}$. Any specific dimensionless data $\overline{\pi}_1$ generates the level surface given by $g(\overline{F}_p(\pi_2,\pi_3,\pi_4))=g(\overline{\pi}_1)$. 

\begin{figure}[htb]
\centering
\subcaptionbox*{$\re{\overline{\pi}_1}=\qty{171.22}{\ohm}$}{\includegraphics[width=.45\linewidth]{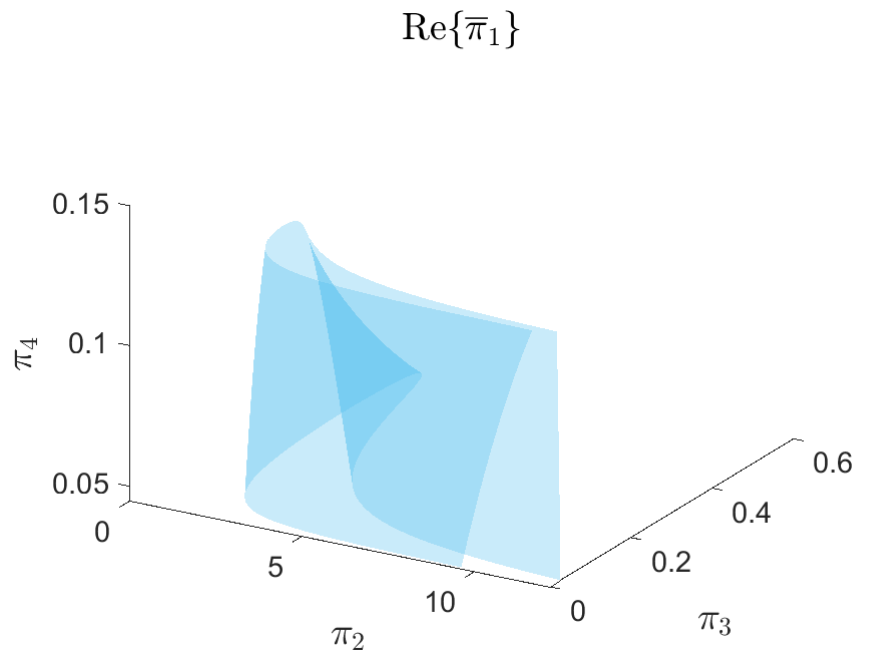}}\hfill 
\subcaptionbox*{$\im{\overline{\pi}_1}=\qty{-332.00}{\ohm}$}{\includegraphics[width=.45\linewidth]{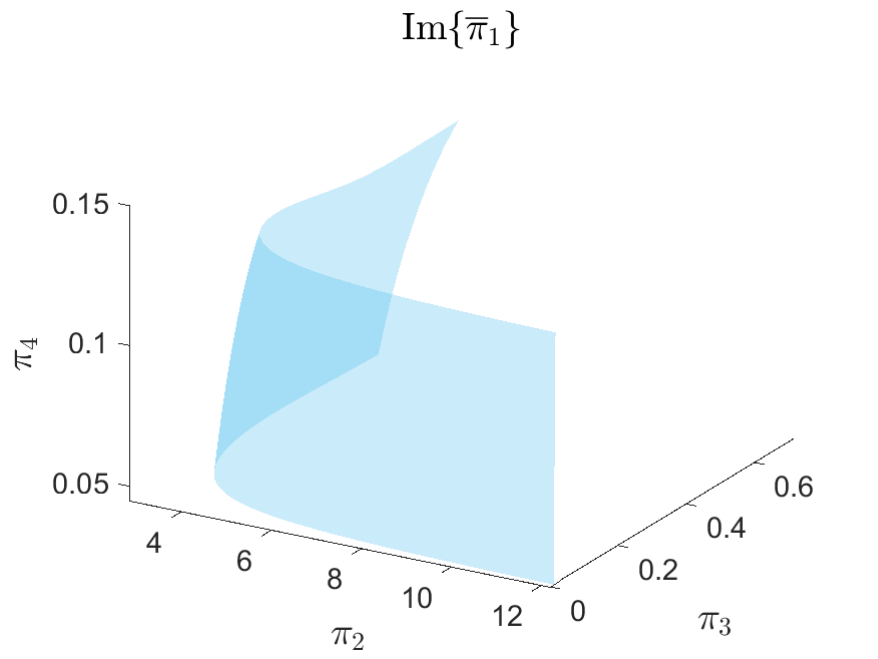}}\hfill
\subcaptionbox*{${|\overline{\pi}_1|}=\qty{373.55}{\ohm}$}{\includegraphics[width=.45\linewidth]{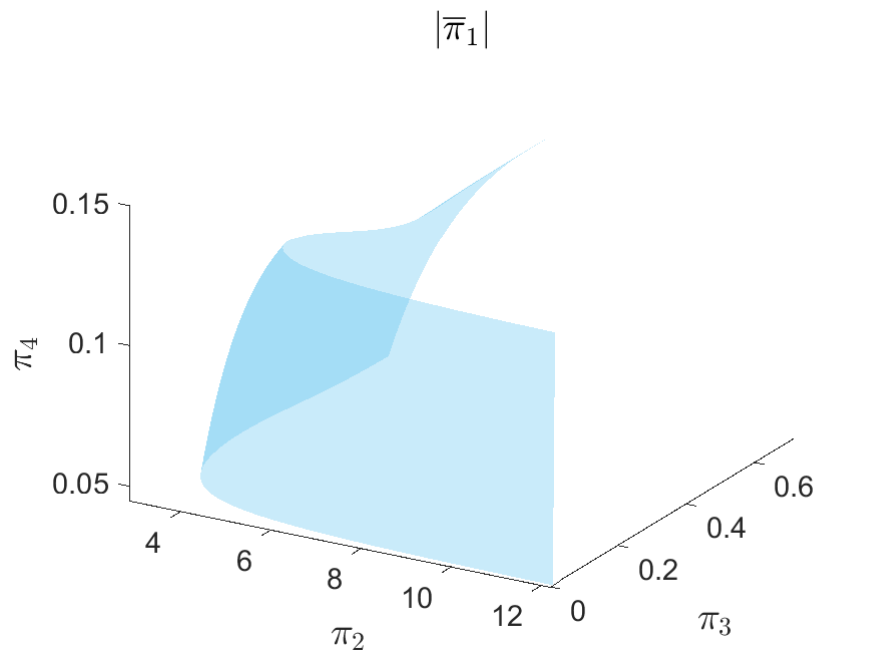}}\hfill
\subcaptionbox*{$\phase{\overline{\pi}_1}=\qty{-1.09}{\radian}$}{\includegraphics[width=.45\linewidth]{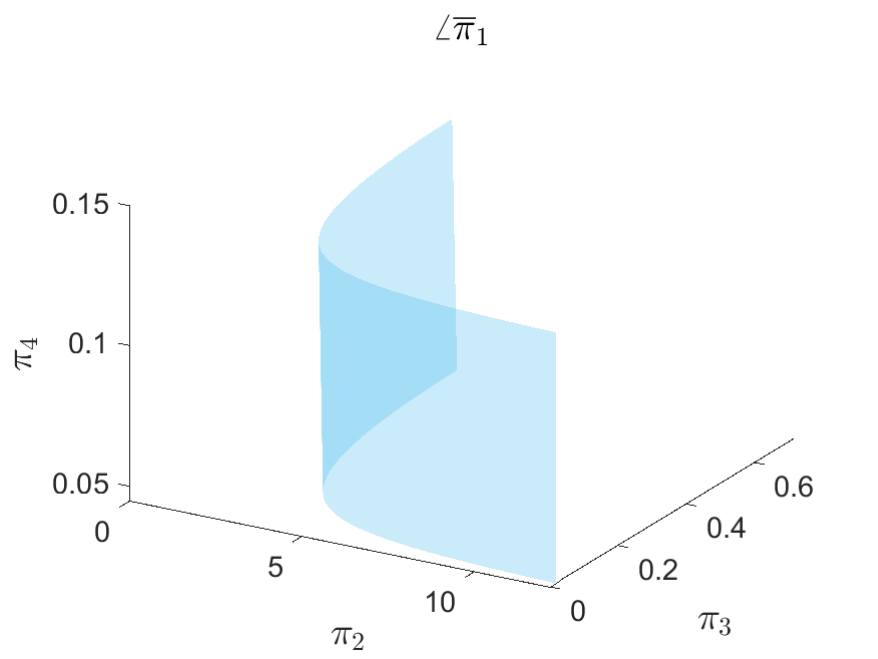}}
\caption{Example of level surface for real part, imaginary part, magnitude and phase. The level is set to $\overline{\pi}_1=\complexqty{171.22 -332.00i}{\ohm}$.}
\label{fig:ls}
\end{figure}

Natural examples of the function $g$ are the real and imaginary parts, or the magnitude and phase of a complex number (see Figure \ref{fig:ls}). For instance, the level surface for $\re{\overline{F}_p(\pi_2,\pi_3,\pi_4)}=\re{\overline{\pi}_1}$ gives the locus of points $(\pi_2,\pi_3,\pi_4) \in \mathbb{P}$ compatible with the real part of the measured data.

\begin{remark}
    The computation of the level surfaces can be easily performed numerically by FEM simulations or employing the analytical/semi-analytical models developed throughout the years for a large variety of ECT problems \cite{Dodd_deeds, An2, An3, An4}. In this specific case the semi-analytic model by Dodd and Deeds \cite{Dodd_deeds} has been employed.
\end{remark}

\begin{remark}
    When processing the measured data, only the intersection of level surfaces is required (see \Cref{level_curves}), which has a negligible computational cost. Moreover, the level surfaces depend only on the prescribed geometry of the probe. Therefore, it is convenient to compute the level surface a priori, for a given probe. 
\end{remark}

\subsubsection{Level curves}
\label{level_curves}
\begin{figure}[htb]
\centering
\includegraphics[width=.9\linewidth]{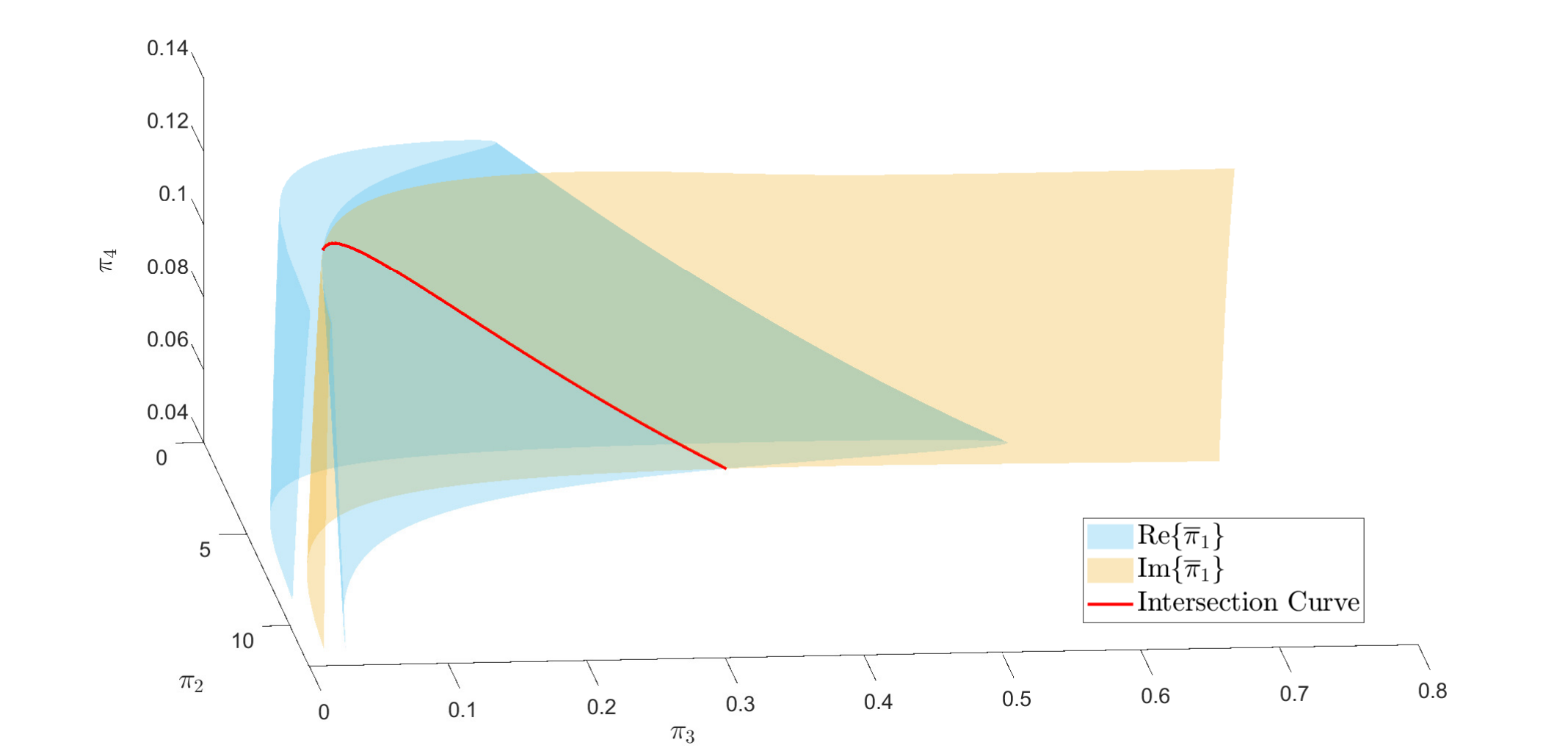}
\caption{Intersection of level surfaces for $\overline{\pi}_1=\overline{\pi}_1^{\omega_1}$.}
\label{fig:intersections}
\end{figure}

Since the dimensionless measured data $\overline{\pi}_1$ is a complex number, it is natural to introduce two families of level surfaces to represent the data, via two functions $g_1,g_2: \mathbb{C} \mapsto \mathbb{R}$. For example, $g_1=\re$ and $g_2=\im$. Therefore, given a measured data $\overline{\pi}_1$, the intersection of the two level surfaces arising from $g_1$ and $g_2$ forms a curve in the 3D parameter space $\mathbb{P}$. This curve is a level curve where both $g_1$ and $g_2$ are constant, and it is described implicitly as
\begin{align} 
g_1(\overline{F}_p(\pi_2,\pi_3,\pi_4)) &= g_1(\overline{\pi}_1) \\ 
g_2(\overline{F}_p(\pi_2,\pi_3,\pi_4)) &= g_2(\overline{\pi}_1)
\end{align}
This curve represents all the possible triples of dimensionless parameters $(\pi_2,\pi_3,\pi_4)$ compatible with the measured data (see Figure \ref{fig:intersections}).

Let a level curve $\gamma$ be given in parametric form as:
\begin{equation}
    \gamma :s\in I\mapsto
     q (s) =
    \left[ 
    \begin{array}{c}
    q _{2}\left( s\right)  \\ 
    q _{3}\left( s\right)  \\ 
    q _{4}\left( s\right) 
    \end{array}
    \right]
    \in \mathbb{P},
\end{equation}
where $I$ is a proper connected interval of $\mathbb{R}$ and $q_2(\cdot)$, $q_3(\cdot)$, and $q_4(\cdot)$ are proper functions.

A relevant property of the level curves is that they are independent of the functions $g_1$ and $g_2$, under weak assumptions. Moreover, they can be characterized via a nonlinear ordinary differential equation (see \Cref{app_A}). In fact, the following proposition holds.
\begin{proposition}
    Under the assumptions of \Cref{app_A}, the evolution of a level curve satisfies a nonlinear ordinary differential equation independent of the functions $g_1$ and $g_2$
    \begin{equation}
        \dot{q} = k(q) \frac{\partial F_p^R}{\partial \Pi} (q) \times \frac{\partial F_p^I}{\partial \Pi} (q),
    \end{equation}
where $\times$ is the cross product, $k(\cdot)$ is a prescribed function with constant sign, and $\partial F_p^R / \partial \Pi$ and $\partial F_p^I / \partial \Pi$ are the gradients of the real and imaginary parts of $\overline{F}_p$ with respect to $\Pi=(\pi_2, \pi_3, \pi_4)$.
\end{proposition}
%\noindent
The function $k(\cdot)$ can be chosen rather arbitrarily. Convenient choices are $k=1$ or $k(q)=\left\|\partial F_p^R  / \partial \Pi \times \partial F_p^I / \partial \Pi \right\|^{-1}$. In the latter case, $s$ corresponds to the curvilinear abscissa. 

The level curves introduced above play a key role in identifying the invariant quantities of interest, as shown in the next sections. In the following, it is convenient to refer to the intersection curves obtained at a prescribed frequency $\omega$ as \emph{compatibility curve} at the frequency $\omega$.

\subsection{Invariants}
\label{sec:invariants}
The idea underlying the systematic computation of the invariants is rather simple from the perspective of the dimensionless parameter space $\mathbb{P}$ and the compatibility curves. Here, the general idea is introduced in terms of dimensionless parameters, while in the following subsections the idea is discussed with reference to the actual physical parameters, where a few specific modifications may be required. 

The procedure for obtaining an invariant transformation, for instance, that with respect to $\pi_4$, is based on the behavior of the compatibility curves when only $\pi_4$ varies. Let $\pi_2^*$ and $\pi_3^*$ be prescribed and let $\pi_4^a$ and $\pi_4^b$ two distinct values for $\pi_4$. Let $\Gamma^a$ and $\Gamma^b$ the compatibility curves (technically the images) crossing the triples $(\pi_2^*,\pi_3^*,\pi_4^a)$ and $(\pi_2^*,\pi_3^*,\pi_4^b)$, i.e. $\Gamma^a$ is generated by the data $\overline{\pi}_1^a=\overline{F}_p(\pi_2^*,\pi_3^*,\pi_4^a)$, while $\Gamma^b$ is generated by the data $\overline{\pi}_1^b=\overline{F}_p(\pi_2^*,\pi_3^*,\pi_4^b)$. In other terms, $\Gamma^a$ is the set of triples $(\pi_2,\pi_3,\pi_4) \in \mathbb{P}$ such that
\begin{align} 
g_1(\overline{F}_p(\pi_2,\pi_3,\pi_4)) &= g_1(\overline{\pi}_1^a) \\ 
g_2(\overline{F}_p(\pi_2,\pi_3,\pi_4)) &= g_2(\overline{\pi}_1^a),
\end{align}
and, similarly, for $\Gamma^b$. By construction $(\pi_2^*,\pi_3^*,\pi_4^a) \in \Gamma^a$ and $(\pi_2^*,\pi_3^*,\pi_4^b) \in \Gamma^b$, therefore, if we project these two curves on the $\pi_2\pi_3-$plane, it results that the $(\pi_2^*,\pi_3^*)$ point belongs to both projections.

Summing up, the following Proposition holds.
\begin{proposition}
    The projections in the $\pi_2\pi_3-$plane of the compatibility curves when $\pi_4$ varies, admit $(\pi_2^*,\pi_3^*)$ as invariant point.
\end{proposition}

The following immediate Corollary holds.
\begin{cor}
    The invariant point of the projection of the compatibility curves in the $\pi_2\pi_3-$plane gives the (dimensionless) values of the parameters of interest.
\end{cor}
    
Therefore, for estimating $(\pi_2^*,\pi_3^*)$ when $\pi_4$ is the (dimensionless) nuisance parameter, it is required to gather the measurements for at least two distinct value of $\pi_4$ and, then, searching for the intersection point for the projection of the compatibility curves in the $\pi_2\pi_3-$plane.

\begin{remark}
    When it is not possible to change parameter $\pi_4$ in at least two different ways, one has to collect and process measurements at different frequencies (see \Cref{sec:ILO}).
\end{remark}

\begin{remark}
\label{rem_multiplePoints}
    When the projections in the $\pi_2\pi_3-$plane of the compatibility curves present multiple intersection points, one has to increase the number of compatibility curves at different $\pi_4$s, and/or use multiple frequencies, and/or change functions $g_1$ and $g_2$.
\end{remark}

This general procedure can be trivially adapted to the cases when $\pi_2$ or $\pi_3$ are the nuisance parameters. In the remaining subsections, this procedure is applied to build the invariant transformations for the lift-off, the electrical conductivity, and the thickness of the non-magnetic plate.

\subsection{Generalization of the Methodology}
The methodology of \Cref{sec:invariants} to generate invariants can easily be extended to (i) problems with more nuisance parameters, and (ii) higher-dimensional problems.

As an example of treatment of the first generalization, let us still consider the model described by \eqref{eqn:amod1}, where both $\pi_3$ and $\pi_4$ are nuisance parameters. The interest is only in estimating $\pi_2^*$.
In this case, given the two measurements $\overline{\pi}_1^a=\overline{F}_p(\pi_2^*,\pi_3^a,\pi_4^a)$ and $\overline{\pi}_1^b=\overline{F}_p(\pi_2^*,\pi_3^b,\pi_4^b)$, it is trivial to prove that the intersection of the projection along the $\pi_2-$axis of the compatibility curves $\Gamma^a$ and $\Gamma^b$ contains at least $\pi_2^*$. In case the intersection of the projections is larger than $\{ \pi_2^* \}$, more data must be collected at different $\pi_3$ and $\pi_4$, as highlighted in \Cref{rem_multiplePoints}.

With reference to the second generalization (higher-dimensional parameter spaces), the extension of the methodology is straightforward from a conceptual point of view. The complexity arises from a practical point of view, since it is required to generate level manifolds in higher dimensions.

Generalization (i) and (ii) can also be combined together, providing a very general systematic method to create invariant transformations.

\subsection{Invariance to lift-off}
\label{sec:ILO}
In this section, the problem of estimating the electrical conductivity and the thickness of a metallic plate is considered, while the lift-off is the nuisance parameter.

\begin{figure}[htb]
\centering
\subcaptionbox*{}{\includegraphics[width=.45\linewidth]{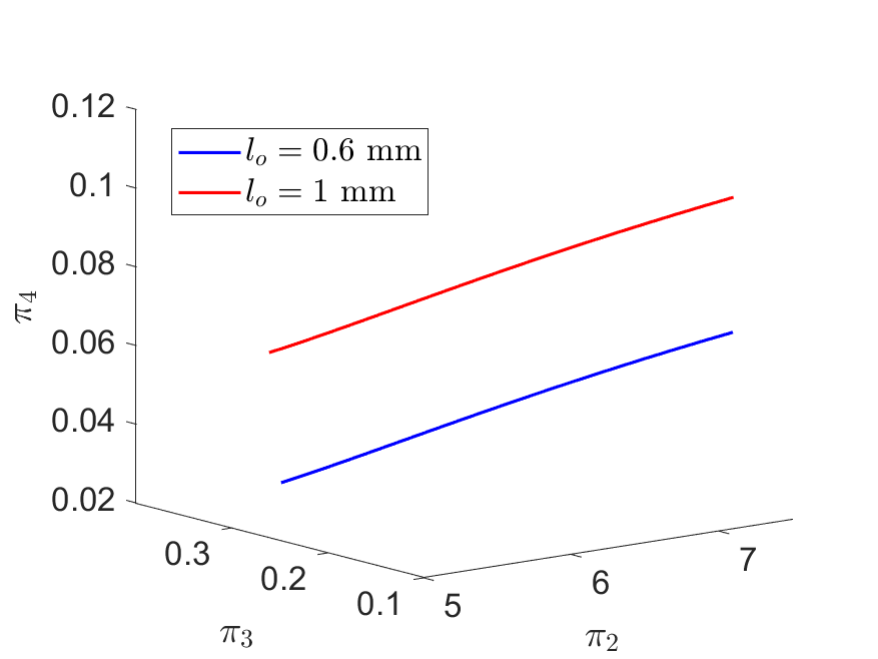}}\hfill 
\subcaptionbox*{}{\includegraphics[width=.45\linewidth]{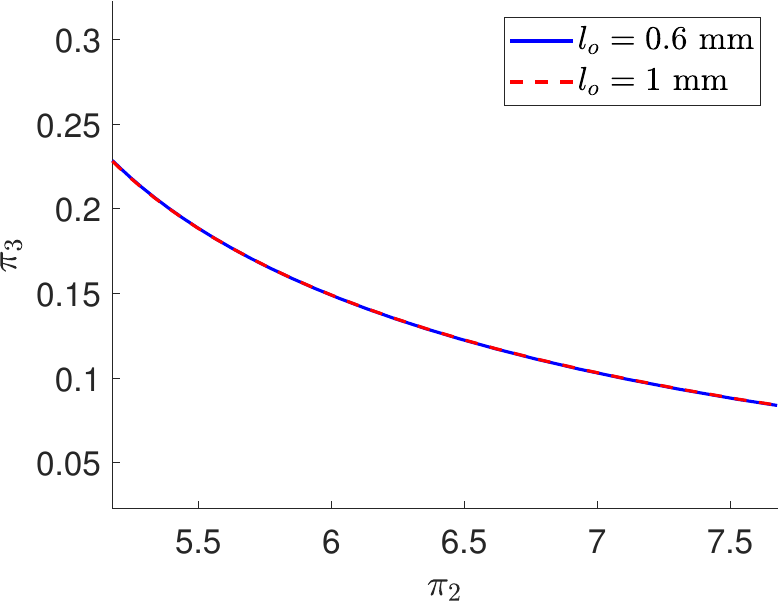}}\hfill
\caption{Two compatibility curves for the same frequency but different lift-off. Left: representation in the dimensionless volume. Right: projection on the $\pi_2 \pi_3-$plane.}
\label{fig:inv_l}
\end{figure}

First, collect two different measurements $\overline{\pi}_1^a$ and $\overline{\pi}_1^b$ at different lift-offs $\pi_4^a$ and $\pi_4^b$, but at the same angular frequency, on a plate with prescribed electrical conductivity and thickness. In line with the method of \Cref{sec:invariants}, the data $\overline{\pi}_1^a$ and $\overline{\pi}_1^b$ generate two different compatibility curves. When projected on the $\pi_2\pi_3-$plane, these two compatibility curves are superimposed (see Figure \ref{fig:inv_l}).
Moreover, an extensive numerical campaign showed that, in general, the compatibility curves at different lift-off have the same projection in the $\pi_2\pi_3-$plane.

%In other words, all the compatibility curves, for the same frequency, share the same projection on the $(\pi_2,\pi_3)$, regardless the lift-off. Thus, the latter projection is \emph{invariant} with respect to the lift-off.

\begin{figure}[htb]
    \centering
    \includegraphics[width=0.5\linewidth]{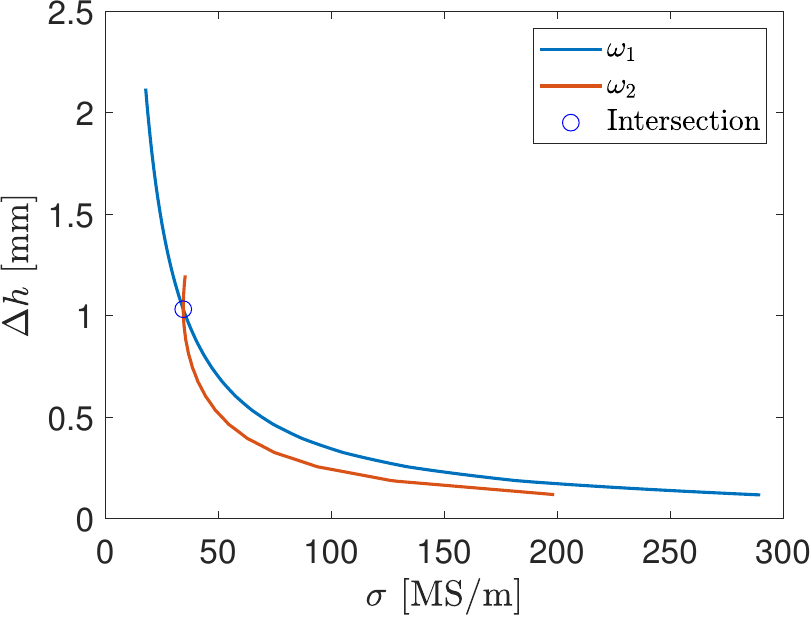}
    \caption{Intersection of two compatibility curves in the $\sigma \Delta h-$ plane, for two different frequencies $f_1=\qty{0.8}{\kilo\hertz}$ and $f_2=\qty{18}{\kilo\hertz}$, and a plate with electrical conductivity \qty{34.5}{\mega\siemens./\meter} and thickness \qty{1.03}{\milli\meter}.}
    \label{fig:inter}
\end{figure}

To obtain the sought estimate for $\sigma$ and $\Delta h$, a second measurement is needed at a different angular frequency. However, this change in the value of the unknown dimensional group $\pi_2$, which depends on the angular frequencies, \lq\lq destroy\rq\rq \ the presence of an invariant point in the projections of the compatibility curves. To restore the presence of an invariant point, the compatibility curves in the $\pi_2\pi_3-$plane are \lq\lq translated\rq\rq \ in terms of curves in the dimensional $\sigma \Delta h-$plane. The intersection of the compatibility curves mapped in this plane provides the estimated values for $\sigma$ and $\Delta h$, as shown in Figure \ref{fig:inter}.

\begin{remark}
    The proposed method allows the simultaneous estimation of electrical conductivity and thickness without the knowledge of the lift-off distance. Furthermore, it is not required to keep the lift-off fixed during the measurements, it can varies with no impact on the estimation.
\end{remark}

Summing up, the proposed method to estimate $\sigma$ and $\Delta h$ without knowledge of the lift-off, requires the following steps:
\begin{enumerate}
    \item collect (at least) two different measurements, at two different frequencies, regardless lift-off;
    \item compute the projection of the compatibility curves in the $\pi_2\pi_3-$plane for the measurements;
    \item translate the compatibility curves from the $\pi_2\pi_3-$plane to the $\sigma \Delta h-$plane. The intersection point in the dimensional plane provides the estimate for $\sigma$ and $\Delta h$.
\end{enumerate}

\subsection{Invariance to electrical conductivity}\label{sec_c_inv}
In this section, a method is proposed to estimate the thickness ($\pi_3^*$) and the lift-off ($\pi_4^*$), without knowing the electrical conductivity ($\pi_2$).

The starting point is given by the compatibility curves corresponding to (at least) two different measurements $\overline{\pi}_1^a$ and $\overline{\pi}_1^b$, in which the dimensional group $\pi_2$ is varied. Since $\pi_2$ depends on the $\omega \sigma$ product, a change in $\pi_2$ can be achieved by a change in the angular frequency, which may be more practical than a change in electrical conductivity.

\begin{figure}[htb]
    \centering
    \includegraphics[width=0.5\linewidth]{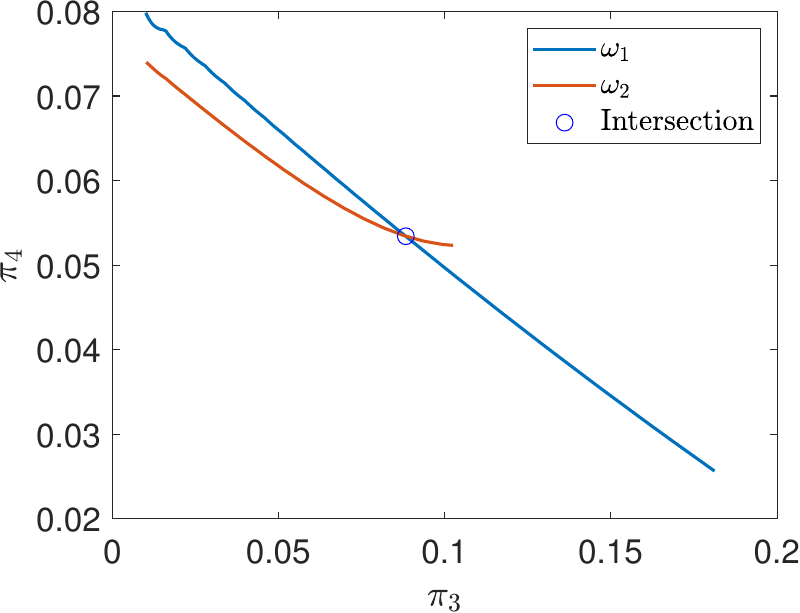}
    \caption{Intersection of two compatibility curves in the $\pi_3 \pi_4-$plane, for two different frequencies $\omega_1=\qty{0.8}{\kilo\hertz}$ and $\omega_2=\qty{18}{\kilo\hertz}$ and thickness \qty{1.03}{\milli\meter} ($\pi_3=8.83\times10^{-2}$) and lift-off distance \qty{0.62}{\milli\meter} ($\pi_4=5.34\times10^{-2}$).}
    \label{fig:inter2}
\end{figure}

The projection of the compatibility curves in the $\pi_3\pi_4-$plane gives the desired $\sigma$ and $\Delta h$ as the intersection of the two projections (see Figure \Cref{fig:inter2}).

The intersection point of the projections does not depend on the current electrical conductivity of the plate, which remains unknown.
Therefore, the method allows one to estimate correctly the thickness and lift-off, even in the presence of variations in the electrical conductivity between different measurements, as in the case of measurements taken at different points of a metallic plate.

\subsection{Invariance to thickness}\label{sec:t_inv}
As a final development, in this section, a method for estimating the electrical conductivity and lift-off of a metallic plate without information about the thickness of the plate is presented. 

Also in this case, the starting point is given by the compatibility curves corresponding to (at least) two different measurements $\overline{\pi}_1^a$ and $\overline{\pi}_1^b$, in which the dimensional group $\pi_3$ is varied. 

\begin{figure}[htb]
    \centering
    \includegraphics[width=0.5\linewidth]{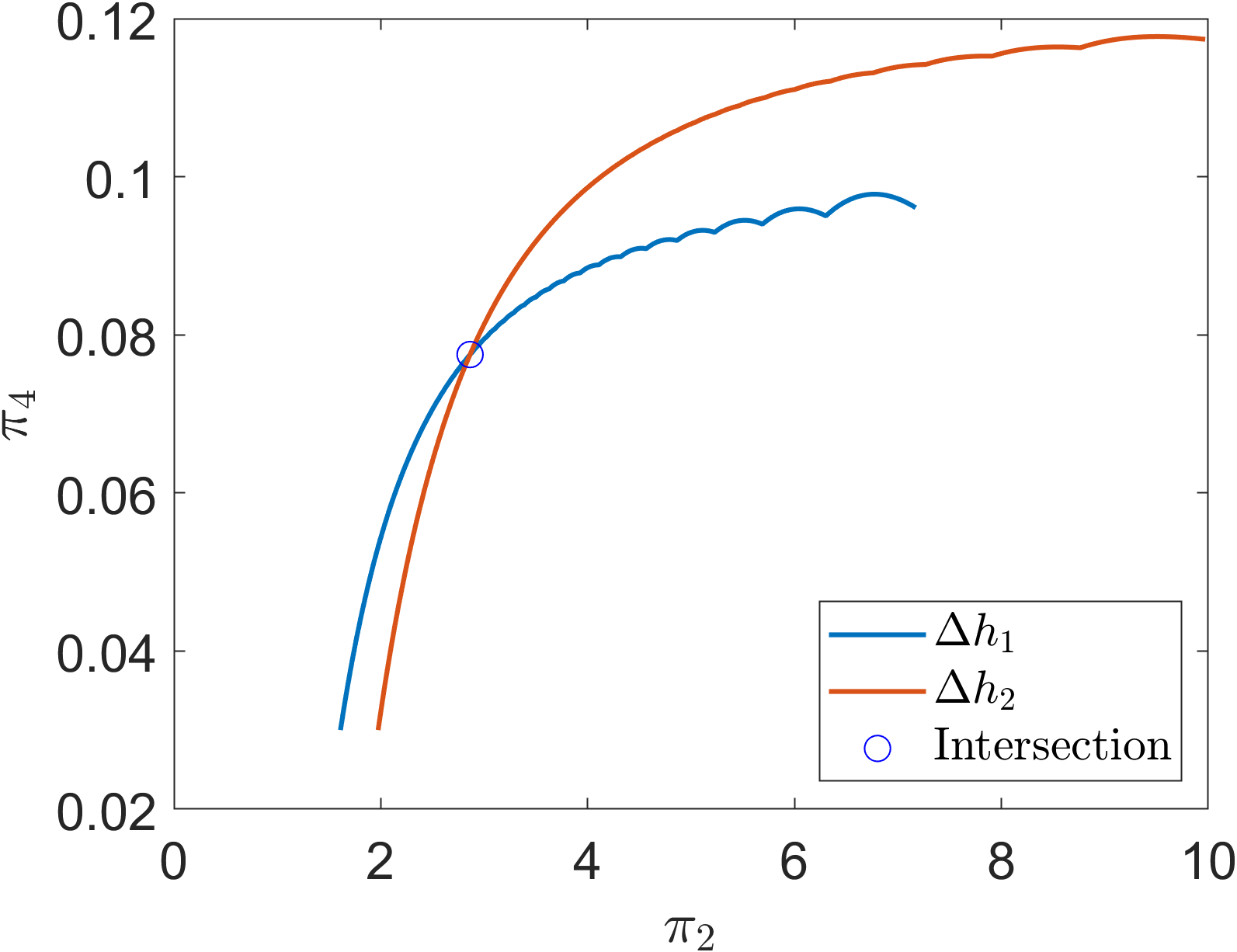}
    \caption{Intersection of two compatibility curves in the $\pi_2\pi_4-$ plane. The compatibility curves are obtained for an electrical conductivity $\sigma=\qty{34.5}{\mega\siemens./\meter}$ and for two different thicknesses $\Delta h_1=\qty{0.5}{\milli\meter}$ and $\Delta h_2=\qty{1}{\milli\meter}$. The measurements are taken at a lift-off distance of $\qty{0.6}{\milli\meter}$ and for a frequency of \qty{1}{\kilo\hertz}. The two curves intersect for $\pi_2=\num{2.86}$ and $\pi_4=7.75\times10^{-2}$, which are the dimensionless values of the unknown parameters.}
    \label{fig:inter3}
\end{figure}

As expected from the theory of \Cref{sec:invariants}, the projection of the compatibility curves $\Gamma^a$ and $\Gamma^b$ in the $\pi_2 \pi_4-$plane gives the electrical conductivity (from $\pi_2$) and the lift-off (from $\pi_4$) via the intersection (invariant) point, as showed in \Cref{fig:inter3}.
The intersection point is \emph{thickness-invariant} and represents the dimensionless counterpart of the unknown electrical conductivity and thickness.

This invariant feature allows for the constructive use of possible irregularities in the plate thickness. If measurements can be taken at different points on the plate, for which a variation in thickness is present, then a simultaneous estimation of the electrical conductivity and lift-off is obtained.

\begin{figure}[htb]
\centering
\subcaptionbox{}{\includegraphics[width=.45\linewidth]{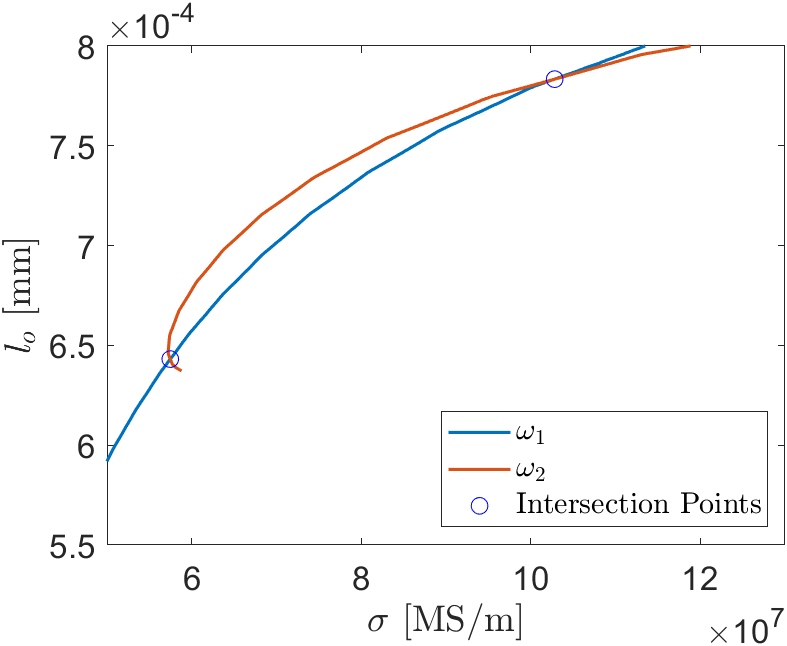}}\hfill 
\subcaptionbox{}{\includegraphics[width=.505\linewidth]{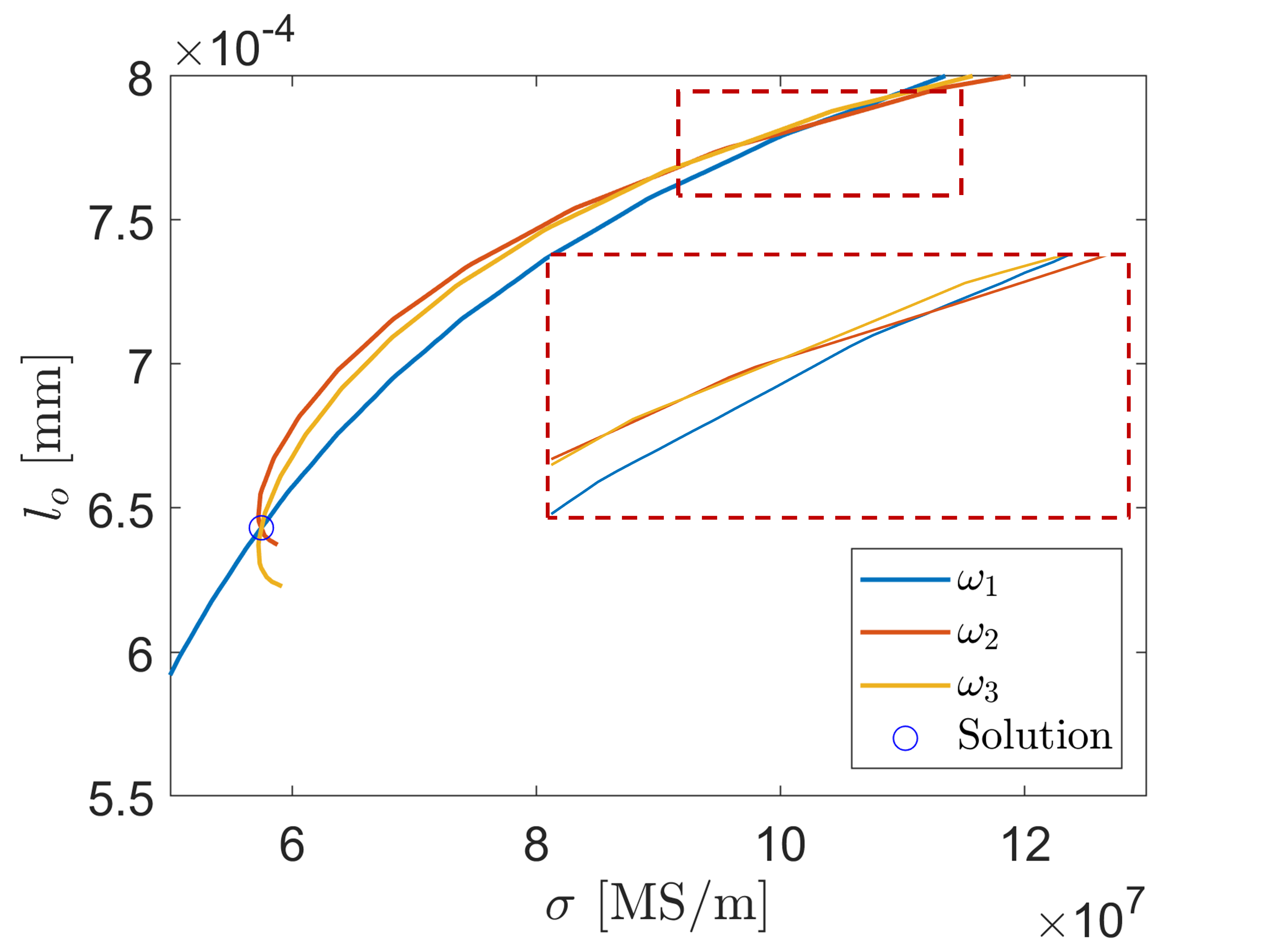}}\hfill
\caption{Two compatibility curves for the same thickness but different frequencies. Left: projections from measurements at two angular frequencies. Right: projections from measurements at three angular frequencies.}
\label{fig:inter4}
\end{figure}

When the thickness of the plate is constant, it is still possible to get the solution ($\sigma$ and $l_o$) via an invariant point, starting from data collected at different frequencies. In this case, to harmonize the compatibility curves, measured at different frequencies, it is necessary to translate the projection of the compatibility curves from the $\pi_2 \pi_4-$plane to the dimensional $\sigma l_o-$plane, as shown in \Cref{fig:inter4}.
This case is interesting because it presents multiple solutions (two solutions) when processing the measurements from two angular frequencies, as shown in \Cref{fig:inter4}(A). This lack of uniqueness can be easily overcome by considering multiple frequencies, as in Figure \ref{fig:inter4}(B). In this case, the solution is given by the only common point shared by \emph{all} the curves.

\section{Experimental validation}\label{sec:exp}
Here, the three variants of the method presented in Section \ref{sec:inter}, to generate invariants, are experimentally validated, testing their accuracy in a realistic environment and different scenarios, where metallic plates with different electrical conductivity and/or thickness are analyzed.

\begin{figure}[htb]
    \centering
    \includegraphics[width=0.95\linewidth]{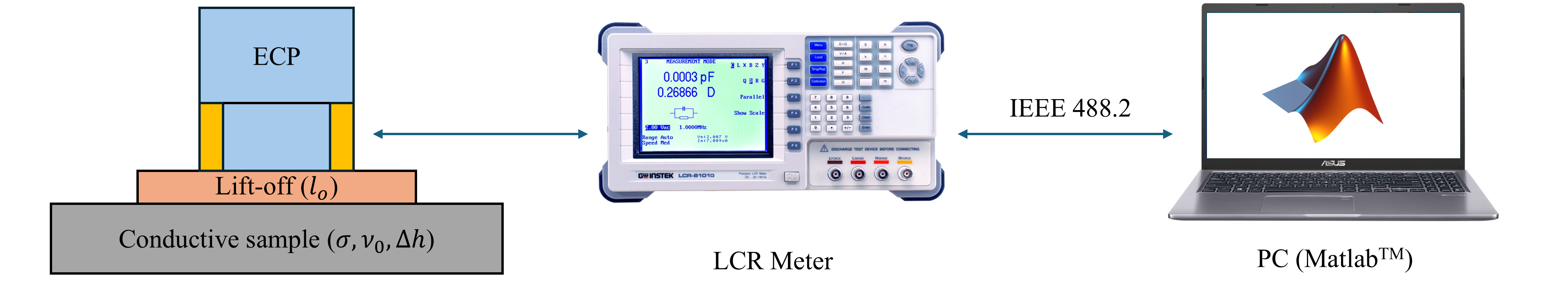}
    \caption{Block diagram of the experimental setup.}
    \label{fig:block}
\end{figure}

The experimental setup is constituted by a remotely controlled LCR meter, an eddy current probe, and a personal computer running in-house software, coded in the Matlab$^{\text{TM}}$ environment, for the acquisition and the signal processing, as shown in Figure \ref{fig:block}.

The eddy current probe (ECP) is an absolute coil, described in \Cref{fig:coils} and Table \ref{tab:coils}.
\begin{table}[htb]
\centering
\captionsetup{justification=centering}
\caption{Characteristics of the ECP employed in the experimental validation.}
\begin{tabular}{cc}
\hline \hline
Parameter  & Value  \\ \hline
$h_c$       & \qty{4.05}{\mm}           \\
$r_i$       & \qty{6.75}{\mm}           \\
$r_e$       & \qty{7.75}{\mm}          \\
$N$         & \num{117}                 \\ 
$\theta$ & \ang{0}             \\ \hline \hline
\label{tab:coils}
\end{tabular}
\end{table}
A GW-Instek 8000G LCR meter is utilized to collect the experimental data, set to slow mode acquisition, and a R-L series model. The meter is connected to a personal computer via the IEEE 488.2 standard interface to perform repeated measurements. 
%Data are acquired at $N=122$ different frequencies, uniformly spaced from $\qty{800}{\hertz}$ to $\qty{25000}{\hertz}$.

Five different plates are considered, where the electrical conductivities and thickness range in a typical interval of interest for industrial applications. The physical properties of the plate are summarized in Table \ref{tab:plates}. Different plastic spacers, whose thicknesses are measured with a digital micrometer, are used to set different lift-off distances between the probe and the specimen. Specifically, the experimental measurements are taken at three different lift-off distances that are $\widetilde{l}_{o,1}=\qty{0.6}{\milli\meter}$, $\widetilde{l}_{o,2}=\qty{1}{\milli\meter}$ and $\widetilde{l}_{o,3}=\qty{1.61}{\milli\meter}$.
\begin{table}[htb]
\captionsetup{justification=centering}
\caption{Metallic plates specifications.}
\begin{tabular}{ccc}
\hline \hline
Name code    & Electrical conductivity ($\widetilde{\sigma}$)\,[\unit{\mega\siemens./\meter}] & Thickness ($\Delta\widetilde{h}$)\,[\unit{\milli\meter}]  \\ \hline
$\#$a      & 28.01  & 1.98  \\
$\#$b      & 35.37  & 1.04  \\
$\#$c      & 35.09  & 1.97   \\
$\#$d      & 34.51  & 2.93    \\  
$\#$e      & 58.05  & 0.99    \\\hline \hline
\label{tab:plates}
\end{tabular}
\end{table}

\subsection{Level surfaces generation}
As discussed in Section \ref{sec:inter}, the first step, common to all the proposed methods, consists of a pre-processing, in which the database for computing $\overline{F}_p$ in \Cref{eqn:amod1} is generated. This database allows to compute the level surfaces and the compatibility curves during the inversion process. The only information needed to build the database is the description of the coil (see Table \ref{tab:coils}) and the range of interest for the unknown parameters.

The database is generated from the numerical evaluation of $\overline{F}_p(\cdot,\cdot,\cdot)$ (see Equation \eqref{eqn:amod1}) by means of the Dodd and Deeds model, where $\pi_2$ ranges in the interval $[0.8,22]$, $\pi_3$ is in the range $[0.01,0.6]$ and $\pi_4$ belongs to $[0.03,0.4]$. These ranges are obtained by running the numerical model for a fixed value $\sigma=\qty{35}{\mega\siemens./\meter}$ and varying the frequency in the range $[34.4,5.83\times 10^4]\,\unit{\hertz}$, the thickness in the range $[0.08,4.6]\,\unit{\milli\meter}$ and the lift-off in the range $[0.23,3.1]\,\unit{\milli\meter}$.

\subsection{Experimental results}
For each plate, lift-off and frequency, $N_m=10$ repeated measurements are collected, to establish the repeatability of the method in estimating the unknown parameters. 

For the selection of operating frequencies, a preliminary numerical analysis was performed to assess the influence of frequency on the estimation performance. The results indicate that, to improve the robustness against measurement noise, the intersection should be determined using two compatibility curves at frequencies $f_i$ and $f_i+\Delta f$, with $\Delta f$ chosen sufficiently large. For this reason, the measurements are taken at two distinct frequency intervals $[800,1600]\,\text{Hz}$ and $[23200,24000]\,\text{Hz}$ and then $N_c=5$ different pairs of compatibility curves are considered. The $i-$th $\mathcal{FP}_i$ consists of two frequencies $f_i$ and $f_i+\Delta f$, where $\Delta f$ is retained constant among all the pairs. The selected frequency pairs $\mathcal{FP}_i$ are summarized in Table \ref{tab:freq}.

\begin{table}[!h]
\captionsetup{justification=centering}
\caption{Frequencies employed in the estimation of the unknown parameters.}
\begin{tabular}{ccccc}
\hline \hline
  $\mathcal{FP}_1$\,[\unit{\kilo\hertz}] & $\mathcal{FP}_2$\,[\unit{\kilo\hertz}] & $\mathcal{FP}_3$\,[\unit{\kilo\hertz}] & $\mathcal{FP}_4$\,[\unit{\kilo\hertz}] & $\mathcal{FP}_5$\,[\unit{\kilo\hertz}] \\ \hline
 \{0.8,23.2\} & \{1.0,23.4\} & \{1.2,23.6\} & \{1.4,23.8\} & \{1.6,24.0\}  \\\hline \hline
\end{tabular}
\label{tab:freq}
\end{table}

In the specific case of estimating electrical conductivity and lift-off via the thickness invariant, a third frequency must be included. Indeed, as pointed out in Section \ref{sec:t_inv}, the estimation can be carried out with two different strategies, either (i) by exploiting differences in the thickness of the plate, or (ii) by employing at least three different frequencies. Since the analyzed plates have constant thickness, the second strategy is implemented. Specifically, a third intermediate frequency $f_t=\qty{12}{\kilo\hertz}$ is considered, to form a frequency triplet $\mathcal{FT}_i=\{f_i,f_t,f_i+\Delta f\}$.

Given a triplet $\mathcal{FT}_i$ for a specific configuration (plate properties plus lift-off), the estimate of the electrical conductivity and lift-off is carried out by considering all the possible intersections between the three compatibility curves and, then, retaining the mean value of each individual result.

Let $\sigma^{i,j}$, $\Delta h^{i,j}$ and $l_o^{i,j}$ be the estimate obtained by considering the $i-$th frequency pair $\mathcal{FP}_i$ or frequency triplet $\mathcal{FT}_i$ (see Tables \ref{tab:freq}) and the $j-$th repeated measurement, the related relative percentage errors are computed as
\begin{equation*}
    \varepsilon_{\sigma}^{i,j}=100\frac{|\sigma^{i,j}-\widetilde{\sigma}|}{\widetilde{\sigma}}, \quad\quad \varepsilon_{\Delta h}^{i,j}=100\frac{|\Delta h^{i,j}-\Delta\widetilde{ h}|}{\Delta\widetilde{ h}}, \quad\quad \varepsilon_{l_o}^{i,j}=100\frac{|l_o^{i,j}-\widetilde{l_o}|}{\widetilde{l_o}}.
\end{equation*}
The mean relative percentage error and the relative error standard deviations are
\begin{itemize}
    \item mean relative percentage error
    \begin{equation*}
        \overline{\varepsilon}_{\sigma}^i=\frac{1}{N_m}\sum_{j=1}^{N_m} \varepsilon_{\sigma}^{i,j},\quad \overline{\varepsilon}_{\Delta h}^i=\frac{1}{N_m}\sum_{j=1}^{N_m} \varepsilon_{\Delta h}^{i,j}, \quad \overline{\varepsilon}_{l_o}^i=\frac{1}{N_m}\sum_{j=1}^{N_m} \varepsilon_{l_o}^{i,j}.
    \end{equation*}
    \item standard deviation of the relative percentage errors
    \begin{equation*}
    \begin{split}
        s_{\sigma}^i&=\sqrt{\frac{1}{N_m-1}\sum_{j=1}^{N_m}\left(\varepsilon_{\sigma}^{i,j}-\overline{\varepsilon}_{\sigma}^i\right)^2}, \\ 
        s_{\Delta h}^i&=\sqrt{\frac{1}{N_m-1}\sum_{j=1}^{N_m}\left(\varepsilon_{\Delta h}^{i,j}-\overline{\varepsilon}_{\Delta h}^i\right)^2}, \\ 
        s_{l_o}^i&=\sqrt{\frac{1}{N_m-1}\sum_{i=1}^{N_m}\left(\varepsilon_{l_o}^{i,j}-\overline{\varepsilon}_{l_o}^i\right)^2}.
    \end{split}
    \end{equation*}
\end{itemize}

\begin{figure}[htb]
\centering
\subcaptionbox*{}{\includegraphics[width=.45\linewidth]{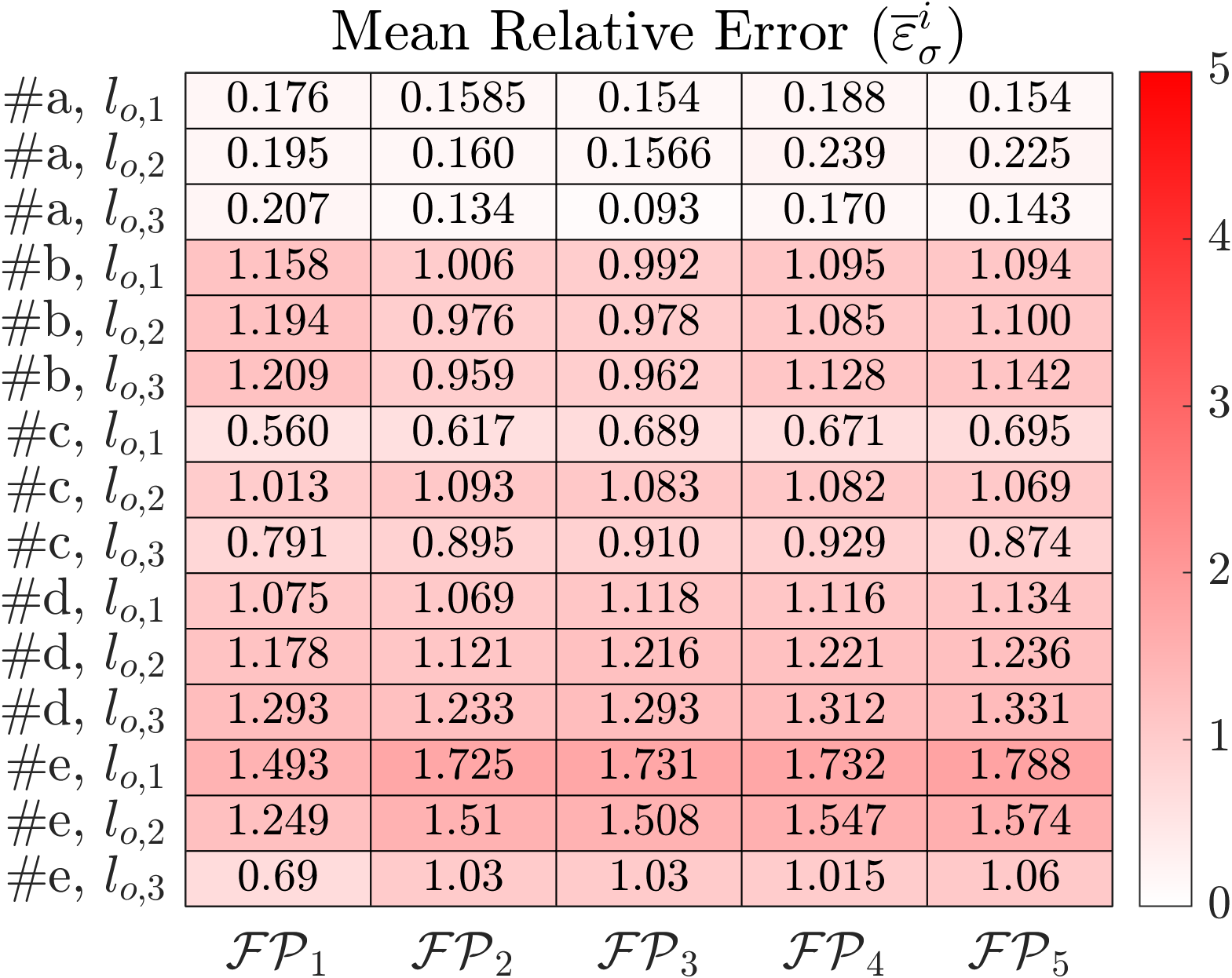}}\hfill 
\subcaptionbox*{}{\includegraphics[width=.45\linewidth]{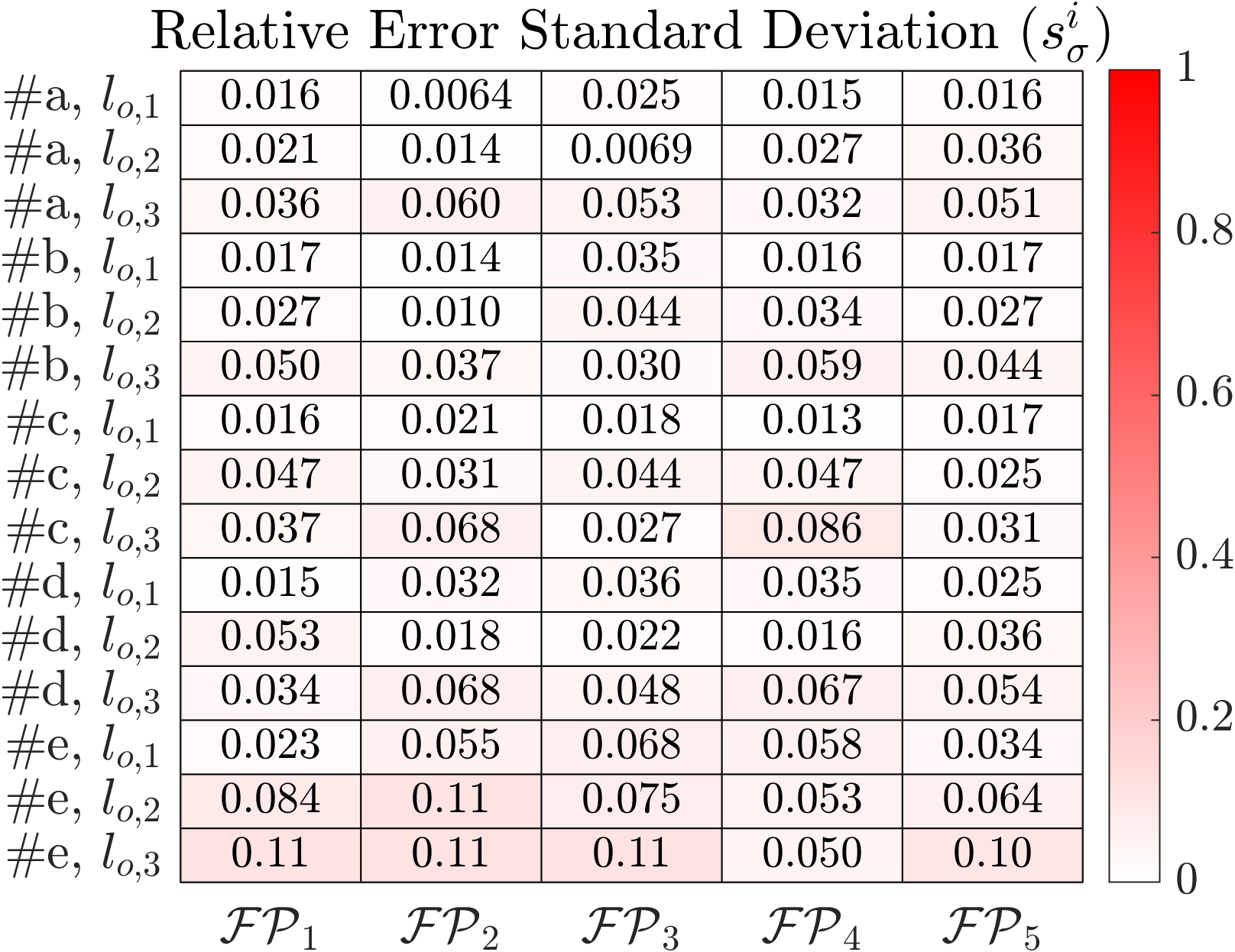}}\hfill
\subcaptionbox*{}{\includegraphics[width=.45\linewidth]{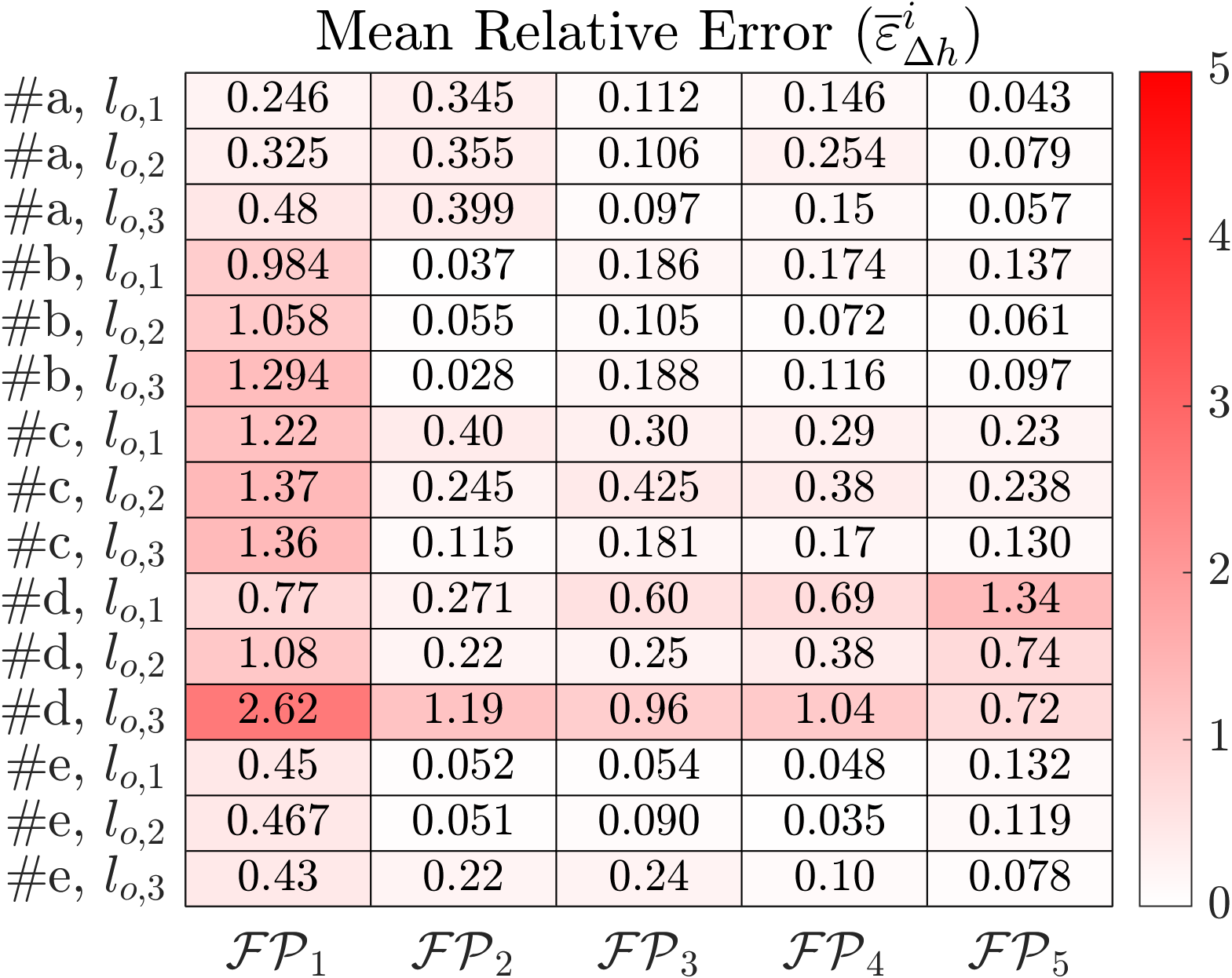}}\hfill
\subcaptionbox*{}{\includegraphics[width=.45\linewidth]{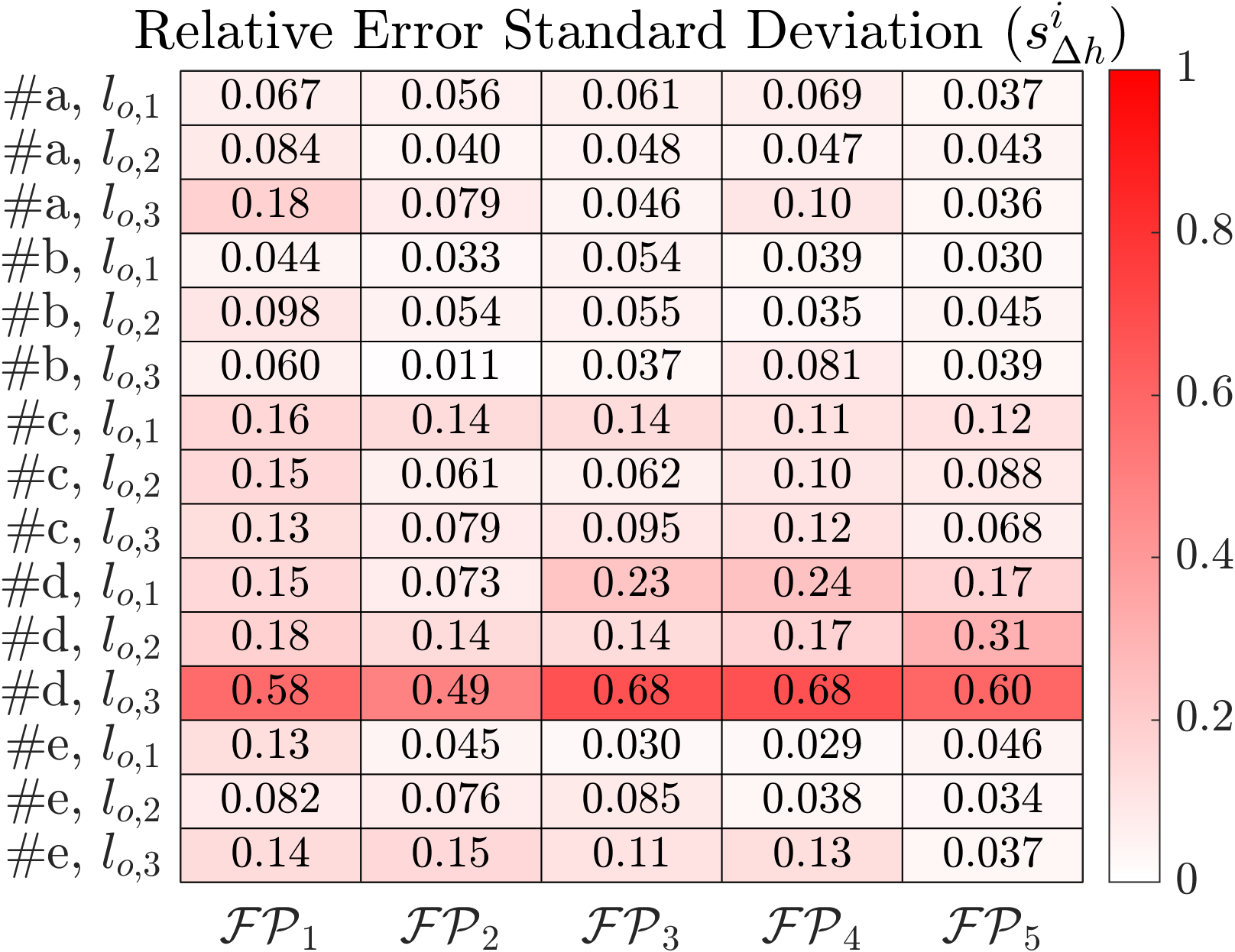}}
\caption{Experimental results for the estimation of electrical conductivity and thickness via lift-off invariant method, for all the combinations of plates, frequencies and lift-offs.}
\label{fig_res_sh}
\end{figure}

\begin{figure}[htb]
\centering
\subcaptionbox*{}{\includegraphics[width=.45\linewidth]{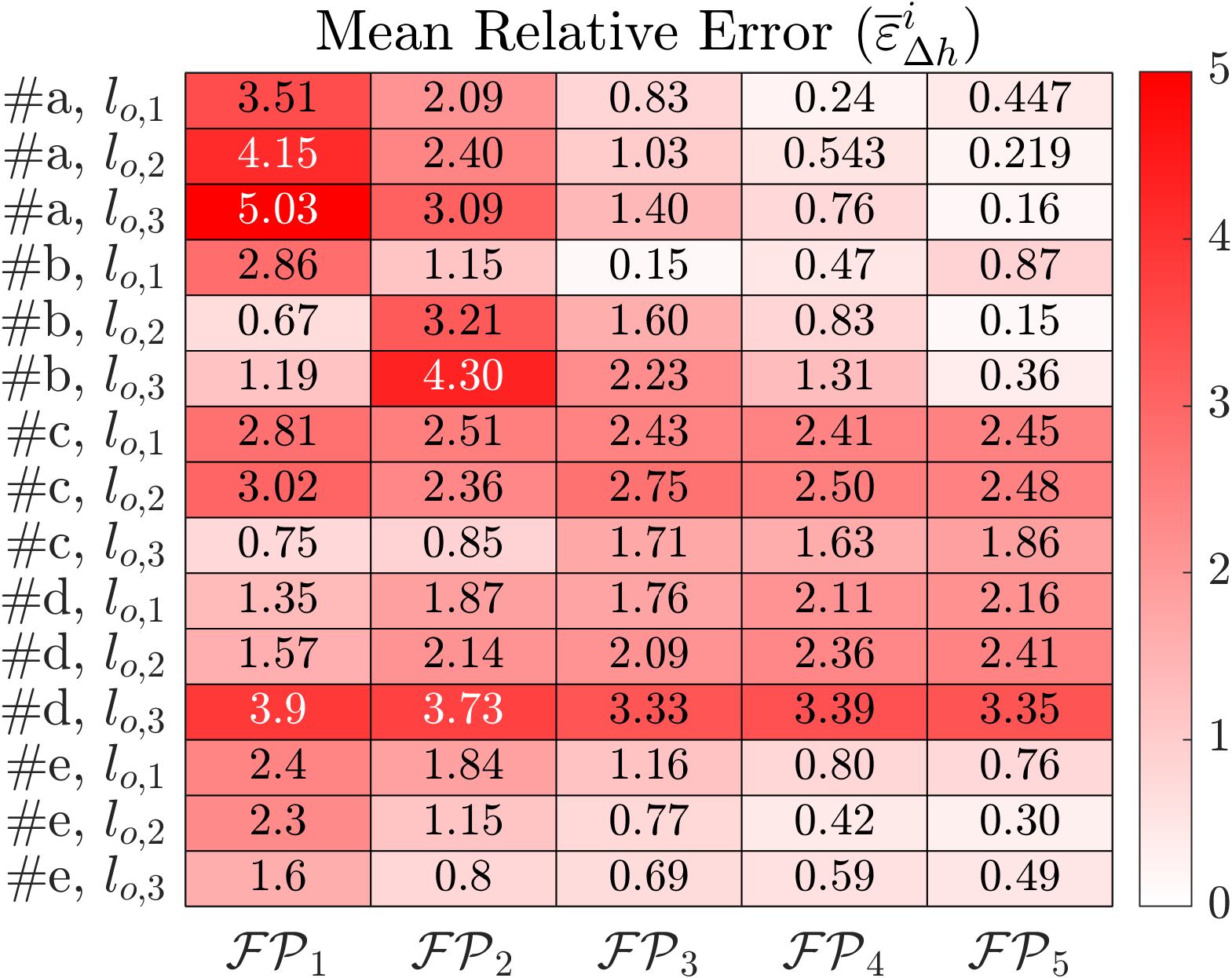}}\hfill 
\subcaptionbox*{}{\includegraphics[width=.45\linewidth]{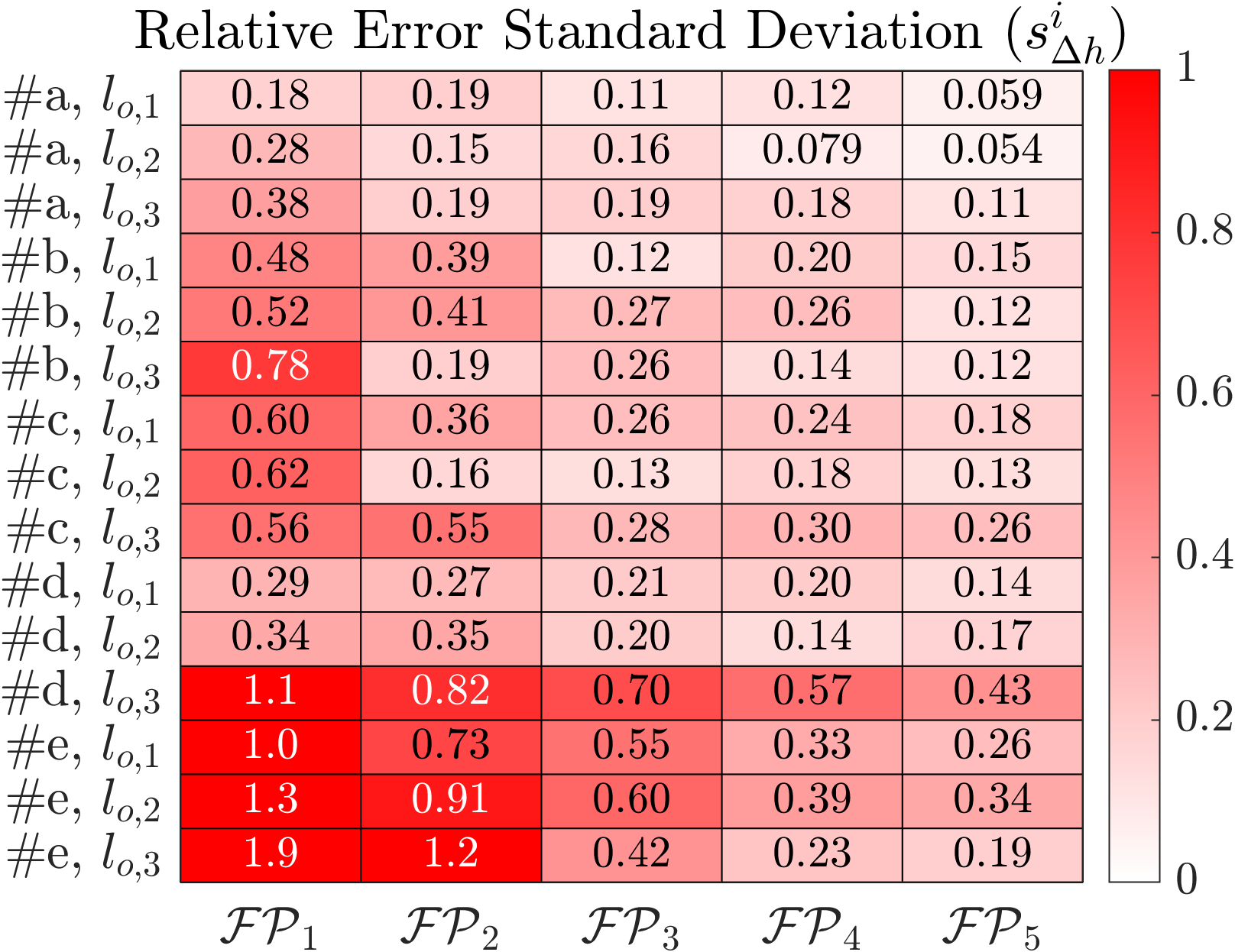}}\hfill
\subcaptionbox*{}{\includegraphics[width=.45\linewidth]{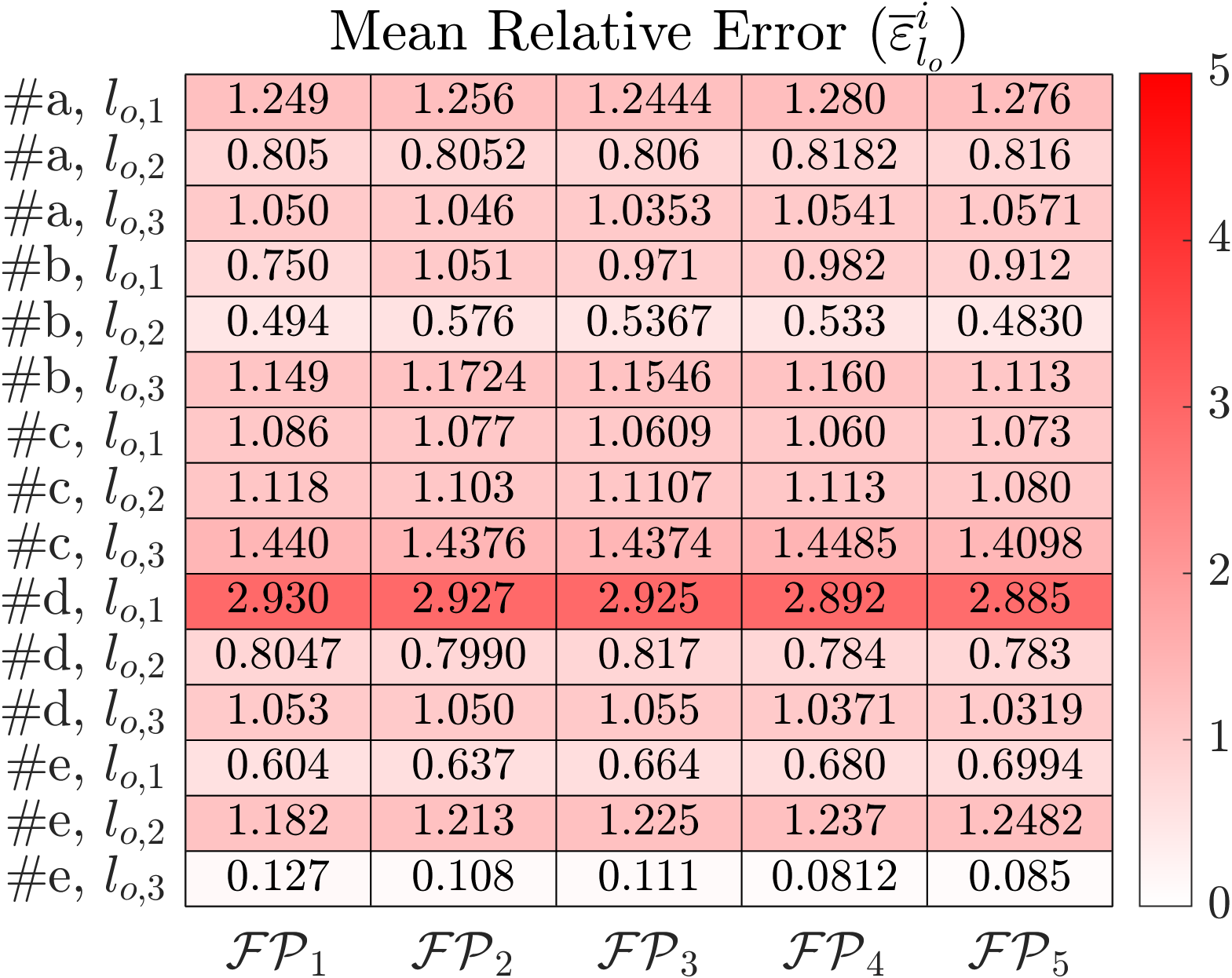}}\hfill
\subcaptionbox*{}{\includegraphics[width=.45\linewidth]{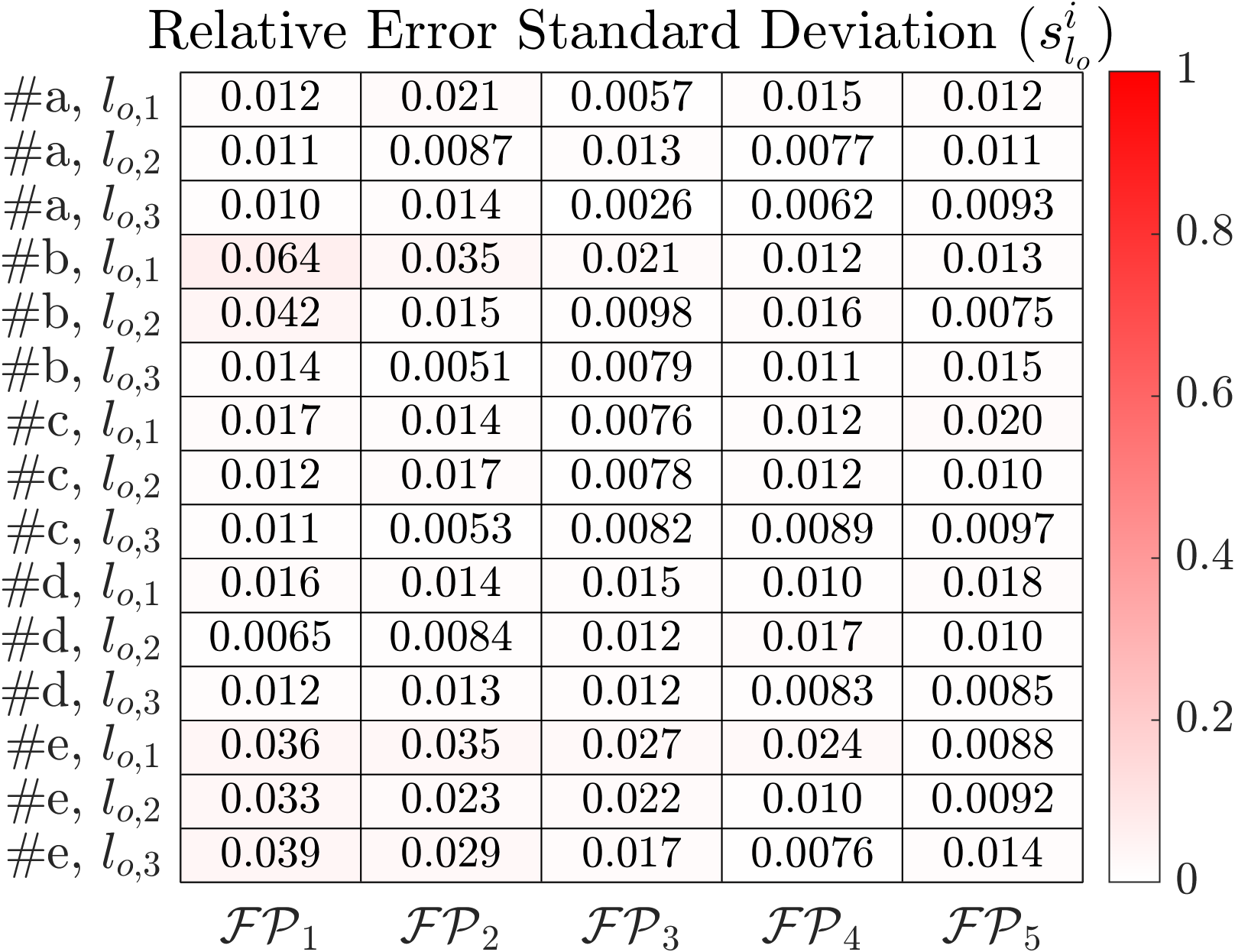}}
\caption{Experimental results for the estimation of thickness and lift-off via electrical conductivity invariant method, for all the combinations of plates, frequencies and lift-offs. }
\label{fig_res_hl}
\end{figure}

\begin{figure}[htb]
\centering
\subcaptionbox*{}{\includegraphics[width=.45\linewidth]{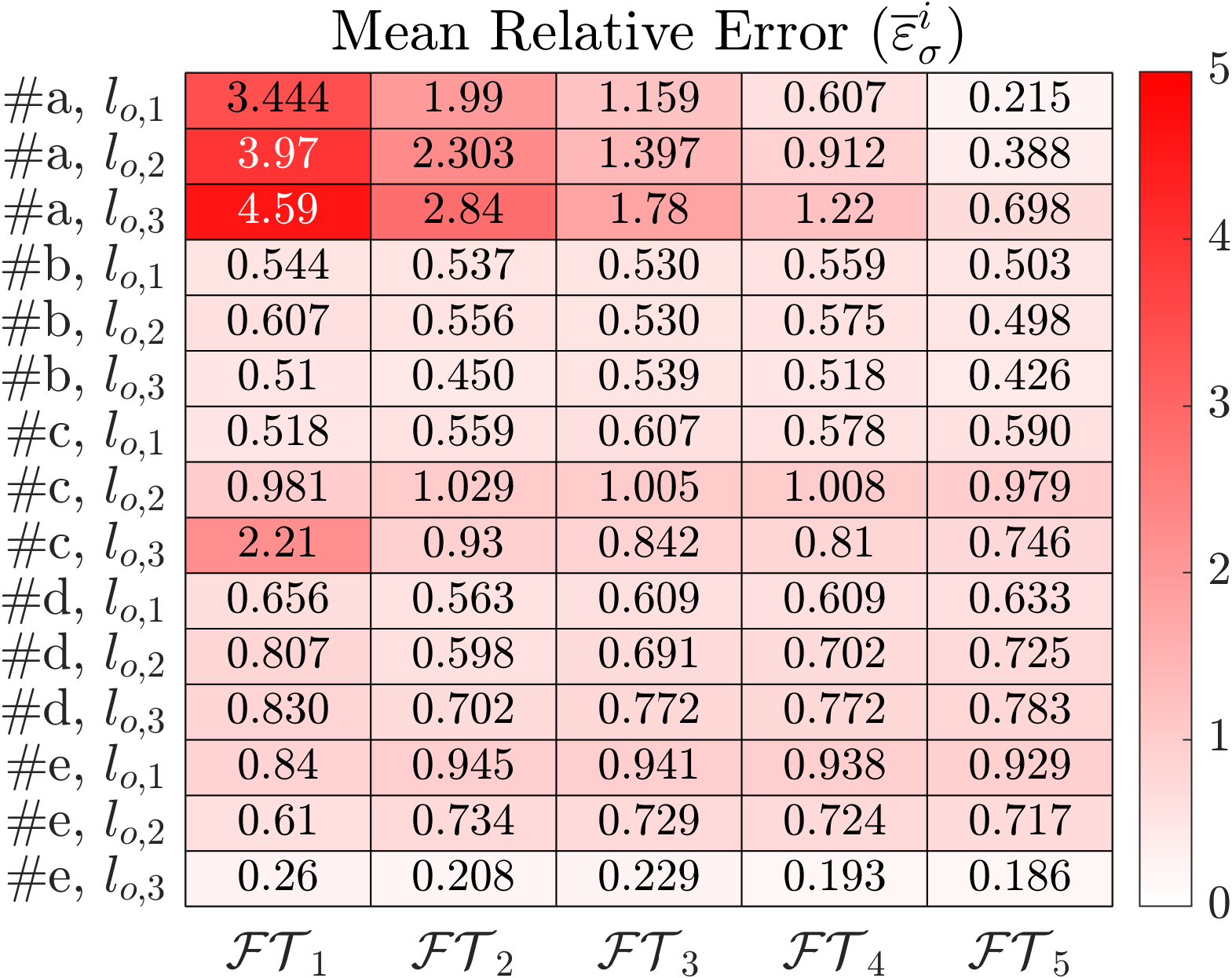}}\hfill 
\subcaptionbox*{}{\includegraphics[width=.45\linewidth]{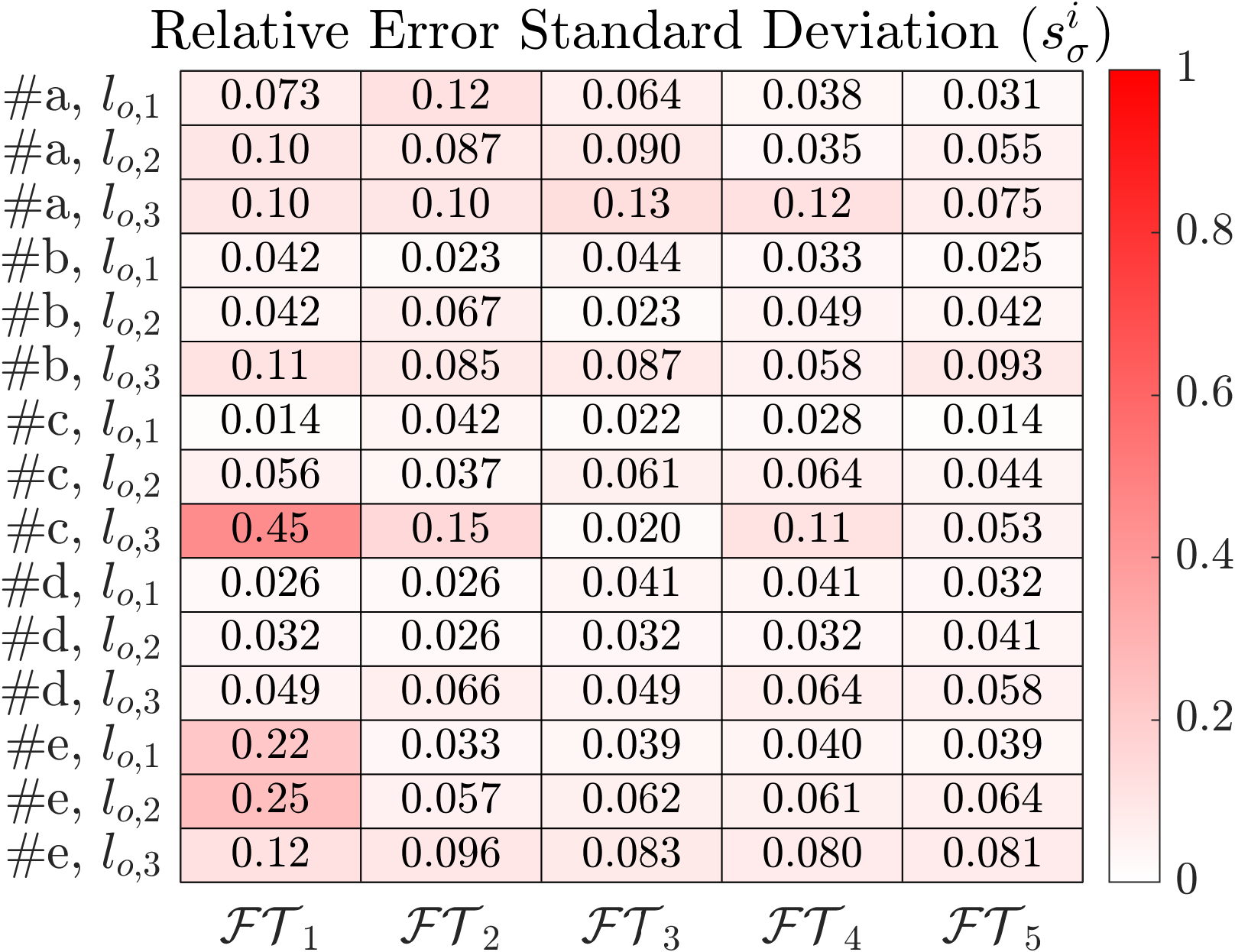}}\hfill
\subcaptionbox*{}{\includegraphics[width=.45\linewidth]{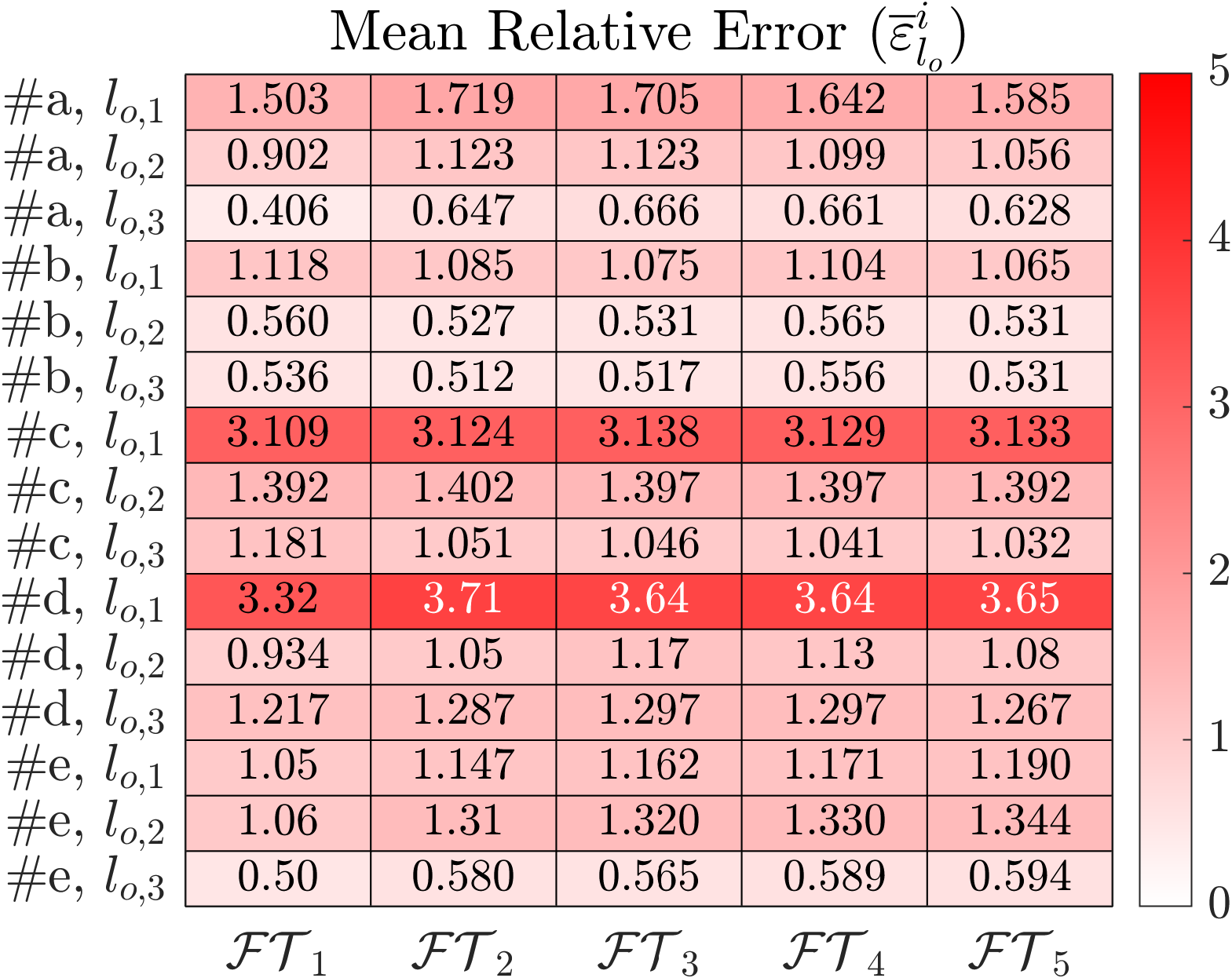}}\hfill
\subcaptionbox*{}{\includegraphics[width=.45\linewidth]{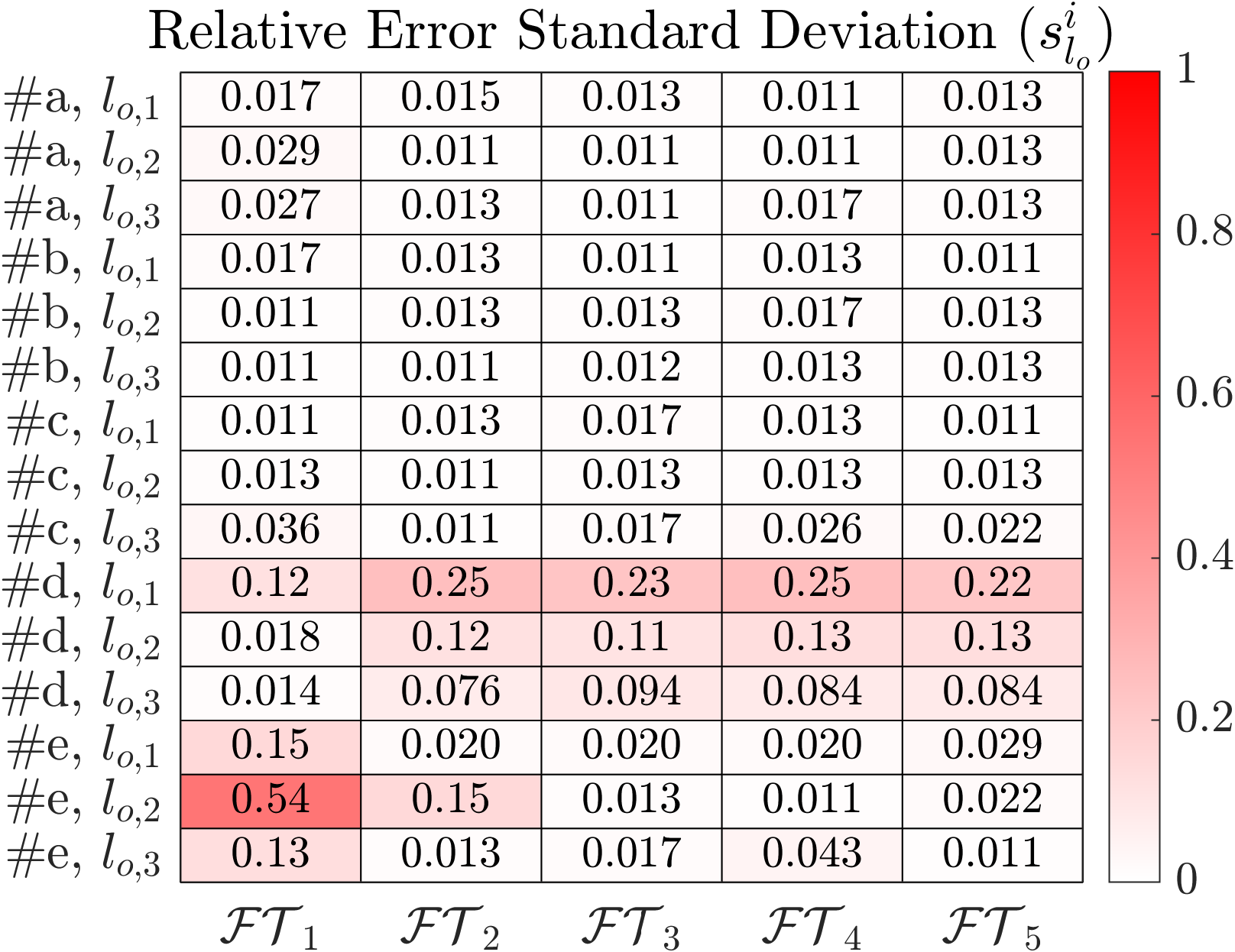}}
\caption{Experimental results for the estimation of the electrical conductivity and lift-off via the thickness invariant method, for all the combinations of plates, frequencies, and lift-offs.}
\label{fig_res_sl}
\end{figure}

The results, shown in Figures \ref{fig_res_sh}, \ref{fig_res_hl}, \ref{fig_res_sl}, clearly demonstrate the effectiveness of the proposed method under all tested conditions and in the analyzed frequency range. Performances in estimating the electrical conductivity and thickness through the invariant lift-off transformation are summarized in Figure \ref{fig_res_sh}. The electrical conductivity is estimated with a mean relative error of less than 1.8\% for all test cases, exhibiting excellent repeatability with a standard deviation of less than 0.10\%. Similar results are obtained for the thickness estimation, with the mean relative error remaining below $1.4\%$, apart from an outlier value at the maximum lift-off distance and the minimum frequency considered for the plate $\# d$. The results related to the estimation of the thickness of the plate $\# d$ at the maximum lift-off distance ($l_{o,3}=\qty{1.61}{\milli\meter}$) are the only ones that show greater variability.

Figure \ref{fig_res_hl} shows the results for the estimation of thickness and lift-off using the $\sigma-$invariant transformation. As it can be observed, the lift-off is estimated with excellent accuracy and repeatability across all the test frequencies and specimens. On the contrary, the thickness estimation is more sensitive to operating frequencies. Despite this, the mean relative error is generally below $3.5\%$ for all test conditions and the standard deviations is generally below $0.6\%$. 

The estimation of electrical conductivity and lift-off using the thickness-invariant transformation demonstrates excellent performance across all frequency-specimen combinations, with mean errors below $2.5\%$ for electrical conductivity and $1.5\%$ for lift-off, as shown in Figure \ref{fig_res_sl}. Only in a few isolated cases performance slightly decreases, yet the resulting errors and standard deviations remain very low.

\section{Conclusion}\label{sec:conclusion}
In this work, a systematic approach to generate invariant transformations for Non destructive Testing \& Evaluation from Eddy Current Testing data has been proposed.

The problem used to present the method is that of the estimation of two parameters, among the electrical conductivity, thickness, and lift-off, for a plate probed by an absolute coil, the third parameter being either uncertain or variable or unknown (nuisance parameter).

A key role is played by Buckingham's \textpi \, theorem, which allows to reduce the number of independent variables required to model the problem.

The approach is based on the projection of the characteristic curves corresponding to the measured data onto a plane perpendicular to the nuisance parameter. For instance, to make a quantity invariant to the lift-off, one has to project the characteristic curves corresponding to the data onto the plane perpendicular to the axis corresponding to the lift-off. 

The method can be extended to problems with a larger number of unknowns or a larger number of nuisance parameters.

An extensive experimental validation confirms the effectiveness of the method in real-world applications, exhibiting errors generally below $3 \%$ in the estimation of the unknown parameters.

\section*{Acknowledgments}
This work was supported by the Italian Ministry of University and Research under the PRIN-2022, Grant Number 2022Y53F3X \lq\lq Inverse Design of High-Performance Large-Scale Metalenses\rq\rq.

\section*{Data availability statement}
The data that support the findings of this study are available from the corresponding author, upon reasonable request.

\appendix

\section{Evolution equation for the level curves}
\label{app_A}
To derive the evolution equation for a level curve, it is convenient to represent $\overline{F}_p$ as a column vector ${F}_p$ consisting of its real and imaginary parts:
\begin{equation}
    {F}_p({\Pi})=\left[ 
    \begin{array}{c}
    F_p^R({\Pi})  \\ 
    F_p^I({\Pi})
    \end{array}
    \right]
\end{equation}
where ${\Pi}$ a point in the parameter space $\mathbb{P}$:
\begin{equation}
    {\Pi} = \left[ 
    \begin{array}{c}
    \pi _{2}  \\ 
    \pi _{3}  \\ 
    \pi _{4}
    \end{array}
    \right].
\end{equation}
Consequently, $g_1$ and $g_2$ are redefined as functions of the plane, i.e. $g_1=g_1(\z)$ and $g_2=g_2(\z)$ where $\z = [z_1,z_2]^T$.

Along a compatibility curve, both $g_1$ and $g_2$ are constants; therefore,
\begin{align} 
    \frac{\text{d}}{\text{d}s} g_1 ( {F}_p (q (s))) &= 0 \\ 
    \frac{\text{d}}{\text{d}s} g_2 ( {F}_p (q (s))) &= 0, 
\end{align}
that gives
\begin{align}
\label{eq_cond}
\diffp{g}{z} ({F}_p ( q (s) )) \  \diffp{{F}_p}{\Pi}(q(s)) \ \dot{q}(s) & = 0,
\end{align}
where the gradients of
$g=\left[
\begin{array}{c}
g_1 \\
g_2 
\end{array}
\right]$ and $F_p$ are:
\begin{align}
    \diffp{g}{z} & =
    \left[
        \begin{array}{cc}
         {\partial g_1}/{\partial z_1}, & {\partial g_1}/{\partial z_2}\\
         {\partial g_2}/{\partial z_1}, & {\partial g_2}/{\partial z_2}\\
        \end{array}
    \right]\\
    \diffp{{F}_p}{\Pi}  & = 
    \left[
    \begin{array}{ccc}
    {\partial F_p^R}/{\partial \pi_2}, &   {\partial F_p^R}/{\partial \pi_3}, & 
    {\partial F_p^R}/{\partial \pi_4} \\
    {\partial F_p^I}/{\partial \pi_2}, &   {\partial F_p^I}/{\partial \pi_3}, & 
    {\partial F_p^I}/{\partial \pi_4} 
    \end{array}
    \right].
\end{align}

If the Jacobian matrix $\partial g / \partial z$ is invertible, then condition \eqref{eq_cond} becomes
\begin{align}
%\label{eq_condApp}
\diffp{{F}_p}{\Pi}(q(s)) \ \dot{q}(s) & = 0,
\end{align}
i.e. $\dot{q}$ must be orthogonal to both $\diffp{{F}_p^R}{\Pi}(q(s))$ and $\diffp{{F}_p^I}{\Pi}(q(s))$. Therefore, $\dot{q}$ is directed along the vector product of $\diffp{{F}_p^R}{\Pi}(q(s))$ and $\diffp{{F}_p^I}{\Pi}(q(s))$:
\begin{equation}
\label{eq_curve}
    \dot{q} = k(q) \diffp{{F}_p^R}{\Pi}(q) \times \diffp{{F}_p^I}{\Pi}(q),
\end{equation}
where $k(\cdot)$ is an arbitrary function.

\begin{remark}
    To obtain equation \eqref{eq_curve} it is required that (i) $F_p$ and $g$ are differentiable and that (ii) $\partial g / \partial z$ is a full rank matrix. 
    Condition (ii), that gives the local invertibility of the $z \mapsto g$ mapping, makes the level curves independent on the choice of functions $g_1$ and $g_2$.
\end{remark}

\bibliographystyle
{plain}
\bibliography{references}
\end{document}